\documentclass[12pt]{article}
\usepackage{a4wide}
\usepackage{graphicx}
\newcommand{\be}{\begin{equation}}
\newcommand{\ee}{\end{equation}}
\newcommand{\ba}{\begin{eqnarray}}
\newcommand{\ea}{\end{eqnarray}}

\begin{document}

\begin{titlepage}
\begin{flushright}
LU TP 07-29\\
%arXiv:yymm.nnnn [hep-ph]\\
Revised November 2007
\end{flushright}
\vfill
\begin{center}
{\Large\bf Isospin breaking in $K\pi$ vector form-factors for the weak and
rare decays $K_{\ell3}$, $K\to\pi\nu\overline\nu$ and $K\to\pi\ell^+\ell^-$}
\vfill
{\bf Johan Bijnens\footnote{Electronic Address: bijnens@thep.lu.se} 
and 
Karim Ghorbani\footnote{Electronic Address: karim.ghorbani@thep.lu.se}  }\\[1cm]
{Department of Theoretical Physics, Lund University,\\
S\"olvegatan 14A, S 223-62 Lund, Sweden}
\end{center}
\vfill
\begin{abstract}
We calculate the two form-factors for the four Kaon to pion transitions via a
vector current to order $p^6$ in Chiral Perturbation Theory to first order
in isospin breaking via the quark masses. In addition we derive relations
between these form-factors valid to first order in the up-down quark-mass
difference but to all orders in Chiral Perturbation Theory.

We present numerical results for all eight form-factors at $t=0$ and for
varying $t$ and for the scalar form-factors at the Callan-Treiman point.
\end{abstract}
\vfill
\vfill
%\footnoterule
%{\footnotesize\noindent }

\end{titlepage}

\section{Introduction}
\label{sect:introduction}

The semileptonic decays of a Kaon to a pion and two leptons play a significant
role in flavour physics. On the one hand, the weak decays $K_{\ell3}$ are
a main source of our knowledge of the CKM matrix element $V_{us}$ and
on the other hand, the rare decays to a lepton-anti-lepton or
neutrino-antineutrino pair provide a good testing bed for loop effects
in flavour physics. The form-factors themselves quantize the hadronic
uncertainties as can be exemplified by the so-called master formulae.
This is e.g. for $K_{\ell3}$, see \cite{Kaon07Cirigliano} and references therein,
\be
\Gamma\left(K^i\to\pi^j\ell^+\nu_l\right)
= C^2_{ij}\frac{G_F^2 S_{EW} m_K^5}{192\pi^3}
\left|V_{us} f_+^{K^i\pi^j}(0)\right|^2{\cal I}^{ij}_\ell
\left(1+2\Delta_{EM}^{ij}\right)\,.
\ee
A similar formula exists for the rare decays, see \cite{Mescia1} and references
therein. We will refer to the $K_{\ell3}$ decays as weak
decays and the ones with a lepton-antilepton pair as rare decays.

Theoretical work on these form-factors goes back a long way.
In Chiral Perturbation Theory (ChPT), the lowest-order (LO) result
dates back to \cite{orderp2x1} while the next-to-leading-order (NLO)
was evaluated by Gasser and Leutwyler \cite{GL2}.
They calculated the vector form-factor $f_+$ for the weak decays
including the isospin breaking
due to $m_u-m_d$ and the scalar form-factor $f_0$ in the isospin limit.
The form-factors are known in the isospin
limit to next-to-next-to-leading order (NNLO) in Chiral Perturbation Theory
(ChPT) \cite{BT3,PS}. In \cite{BT3} a comparison with experimental results
was done and a useful relation for the order $p^6$ constants needed for
$f_+(0)$ obtained. The isospin breaking to NLO for
the vector form-factors $f_+(t)$ for the rare decays was done in \cite{Mescia1}
to NLO. The electromagnetic corrections to NLO, i.e. order $e^2p^2$, are
also known, for the weak decays \cite{Cirigliano0}
and for the rare decays \cite{Mescia1}.
In this paper we calculate the isospin breaking corrections 
due to the quark-mass difference $m_u-m_d$
to the vector
and scalar form-factors to NNLO order in ChPT for all eight form-factors.
The NNLO results are new for all form-factors while the 
NLO results are new for the scalar form-factors.
Some preliminary results
were reported in \cite{Kaon07}.
In addition we discuss the results on ratios of form-factors to NNLO.
Some of these ratios were observed to have special features at NLO
in \cite{GL2} and \cite{Mescia1}. We prove that the relations
(\ref{reliso1}) and (\ref{reliso2}) are valid to all
orders in ChPT to first order in $m_u-m_d$. The double ratio (\ref{reliso2})
was also
discussed in \cite{Mescia1} but not proven there.
There exists also work using dispersion relation for the form-factors
in the isospin limit, see \cite{Passemar,Pichetal} and references therein.

This paper is organized as follows. We define the form-factors
 and derive the relations the form-factors
should satisfy to all orders in ChPT and first order in $m_u-m_d$
in Sect.~\ref{sect:amplitude}.
Next we give a short discussion of ChPT in Sect.~\ref{sect:chpt}
and derive how $\pi^0$-$\eta$ mixing can be taken into account to NNLO
in ChPT in Sect.~\ref{sect:mixing}. Sect.~\ref{sect:ratios} defines the various
ratios of form-factors we use and discusses
how they are obeyed at NLO and NNLO.
We also discuss there some general aspects of our calculation
and give the LO results.
Explicit formulas are not provided at NNLO, they are simply too long
but we present the NLO formulas in App.~\ref{App:NLO} and the dependence
on the order
$p^6$ low-energy constants (LECs) in App.~\ref{App:Ci}. The estimate
of these order $p^6$ LECs we use is presented in Sect.~\ref{sect:resonance}.
Our main results are presented numerically in Sect.~\ref{sect:numerics}.
These include, numerical results on the values of $f_+(0)$, its $t$-dependence,
ratios as a function of $t$ and the deviation from $F_K/F_\pi$ at the
Callan-Treiman point. A short summary is given in Sect.~\ref{sect:conclusions}.

\section{Form-factors and isospin relations}
\label{sect:amplitude}

In this paper we deal with the four matrix-elements
\ba
\label{def+0}
\langle \pi^0 (p') | \overline s\gamma_\mu u(0)| K^+(p)\rangle
&=& \frac{1}{\sqrt{2}} \left[(p'+p)_\mu f^{K^+\pi^0}_+ (t) + (p-p')_\mu
f_-^{K^+\pi^0} (t)\right]\,,
\\
\label{def0-}
\langle \pi^- (p') | \overline s\gamma_\mu u(0)| K^0(p)\rangle
&=& \left[(p'+p)_\mu f^{K^0\pi^-}_+ (t) + (p-p')_\mu
f_-^{K^0\pi^-} (t)\right]\,,
\\
\label{def++}
\langle \pi^+ (p') | \overline s\gamma_\mu d(0)| K^+(p)\rangle
&=& \left[(p'+p)_\mu f^{K^+\pi^+}_+ (t) + (p-p')_\mu
f_-^{K^+\pi^+} (t)\right].
\\
\label{def00}
\langle \pi^0 (p') | \overline s\gamma_\mu d(0)| K^0(p)\rangle
&=& \frac{-1}{\sqrt{2}} \left[(p'+p)_\mu f^{K^0\pi^0}_+ (t) + (p-p')_\mu
f_-^{K^0\pi^0} (t)\right].
\ea
We have thus in total a set of 8 form-factors. They depend
on 
\be
t = (p'-p)^2\,,
\label{s36}
\ee
the square of the four momentum transfer to the leptons.
The form-factors are normalized such that
\be
f_+^{K^i\pi^j}(0) = 1
\ee
in the $SU(3)$ limit of $m_u=m_d=m_s$.
In the isospin limit
\be
f_\pm = f_\pm^{K\pi}=f_\pm^{K^+\pi^0} = f_\pm^{K^0\pi^-}
=f_\pm^{K^+\pi^+} = f_\pm^{K^0\pi^0}\,.
\ee
$f_+^{K\pi}$ is referred to as the vector form-factor, because
it specifies the $P$-wave  projection of the crossed channel matrix-elements
 $\langle\overline s\gamma_\mu q(0)| \mid K^i, \pi^j \;\mbox{in} >$.
 The $S$-wave projection is described by the scalar form-factor
\be
f^{K^i\pi^j}_0 (t) = f^{K^i\pi^j}_+ (t) + \frac{t}{m^2_{K^i} - m^2_{\pi^j}}
 f^{K^i\pi^j}_-(t)
\,.
 \label{deff0}
\ee
We will refer to the decays as the charged weak for (\ref{def+0}),
neutral weak for (\ref{def0-}), charged rare for (\ref{def++})
and neutral rare for (\ref{def00}).

In this paper we derive the isospin breaking due to the quark-mass difference
$m_u-m_d$ to NNLO for the eight form-factors defined above.
We do this to first order in isospin breaking. Let us now derive
first some general properties.
The isospin-breaking
operator $(1/2)\left(m_u-m_d\right)\left(\overline uu-\overline dd\right)$
has isospin one. The pions have isospin one and the Kaons as well as the vector
operator are in an isospin $1/2$ multiplet.
To first order in isospin breaking from $\delta=m_u-m_d$
the form-factors described above can be rewritten in the form
\ba
\label{expansion}
f_\ell^{K^+\pi^0}(t) &=& f_\ell^A(t) +\delta f_\ell^B(t)+{\cal O}(\delta^2)\,,
\nonumber\\
f_\ell^{K^0\pi^-}(t) &=& f_\ell^A(t) -\delta f_\ell^D(t)+{\cal O}(\delta^2)\,,
\nonumber\\
f_\ell^{K^+\pi^+}(t) &=& f_\ell^A(t) +\delta f_\ell^D(t)+{\cal O}(\delta^2)\,,
\nonumber\\
f_\ell^{K^0\pi^0}(t) &=& f_\ell^A(t) -\delta f_\ell^B(t)+{\cal O}(\delta^2)\,,
\ea
for $\ell=+,-,0$. The form (\ref{expansion}) is a direct consequence of the
Wigner-Eckart theorem. This can be interpreted as that the size of isospin
breaking depends on the final pion and the sign also depends
on which kaon is in the initial state.

As a consequence of (\ref{expansion}) we obtain the relations
\be
\label{reliso1}
f_\ell^{K^+\pi^0}(t)-f_\ell^{K^0\pi^-}(t)-f_\ell^{K^+\pi^+}(t)
+f_\ell^{K^0\pi^0}(t)=0+{\cal O}(\delta^2)\,,
\ee
and
\be
\label{reliso2}
r(t)\equiv \frac{f_\ell^{K^+\pi^0}(t) f_\ell^{K^0\pi^0}(t)}
     {f_\ell^{K^0\pi^-}(t)f_\ell^{K^+\pi^+}(t)} = 1+
{\cal O}(\delta^2)
\ee
These relations do not have to be satisfied when electromagnetic corrections
are included. Photon exchange contains isospin 0, 1 and 2 parts allowing
different corrections to all four amplitudes. The isospin 0 and 1 parts
do satisfy the same relations, but not the isospin 2 part.

The relations are valid for all three form-factors $f_\ell^{K^i\pi^j}$ with
$\ell=+,-,0$. They are also valid if the currents in (\ref{def+0}-\ref{def00})
are replaced by the scalar densities $\overline su(0)$ and $\overline s d(0)$.
 
\section{Chiral Perturbation Theory}
\label{sect:chpt}

Chiral Perturbation Theory (ChPT) is an
effective field theory to describe the strong 
interactions at very low energy. The effective Lagrangian 
is constructed based on two important properties of the
physical hadron spectrum. Pseudo-scalar mesons, the lowest-lying 
states in the spectrum are separated from the rest of the
hadrons, i.e. there exists a mass gap. 
This allows the heavier particles to decouple
from the dynamics of the pseudo-scalar mesons. Their influence
can be described by point-like couplings.
The other important fact is that the spectrum does
not show the chiral symmetry of the underlying theory (QCD).
The pseudo-scalars are assumed to be the
the pseudo-Goldstone particles emerging from the spontaneous 
breaking of this chiral symmetry.
The nonzero but
small mass of the pseudo-scalar mesons are because quarks have a finite mass 
in, reality which breaks the chiral symmetry explicitly.
 
According to the Goldstone's theorem, the Goldstone particles 
do not interact at zero momentum. This immediately offers a
weakly interacting theory as a basis for perturbation theory.
The first systematic consideration on the applicability 
of the effective Lagrangians was made by Weinberg
\cite{Weinberg0} and Gasser and Leutwyler \cite{GL0}.
The effective chiral Lagrangian is an expansion in momentum
and quark masses. In the chiral power-counting, quark masses are 
of order $p^2$. Taking into account the Lorentz invariance and chiral symmetry,
the lowest order chiral Lagrangian which also complies with the discrete 
symmetries can be written down as 
\be
\mathcal{L}_{2} = \frac{F_0^{2}}{4} \left( \langle D_{\mu}U D^{\mu}U^{\dagger}
 \rangle + 
\langle\chi U^{\dagger} + U \chi^{\dagger}  \rangle  \right) 
\ee
and the next-to-leading Lagrangian with the introduction of the 
external field technique was written down by Gasser and Leutwyler
\cite{GL1} 
and reads 
\ba
\label{lagL4}
{\cal L}_4&&\hspace{-0.5cm} = 
L_1 \langle D_\mu U^\dagger D^\mu U \rangle^2
+L_2 \langle D_\mu U^\dagger D_\nu U \rangle 
     \langle D^\mu U^\dagger D^\nu U \rangle \nonumber\\&&\hspace{-0.5cm}
+L_3 \langle D^\mu U^\dagger D_\mu U D^\nu U^\dagger D_\nu U\rangle
+L_4 \langle D^\mu U^\dagger D_\mu U \rangle
 \langle \chi^\dagger U+\chi U^\dagger \rangle
\nonumber\\&&\hspace{-0.5cm}
+L_5 \langle D^\mu U^\dagger D_\mu U (\chi^\dagger U+U^\dagger \chi ) \rangle
+L_6 \langle \chi^\dagger U+\chi U^\dagger \rangle^2
\nonumber\\&&\hspace{-0.5cm}
+L_7 \langle \chi^\dagger U-\chi U^\dagger \rangle^2
+L_8 \langle \chi^\dagger U \chi^\dagger U
 + \chi U^\dagger \chi U^\dagger \rangle
\nonumber\\&&\hspace{-0.5cm}
-i L_9 \langle F^R_{\mu\nu} D^\mu U D^\nu U^\dagger +
               F^L_{\mu\nu} D^\mu U^\dagger D^\nu U \rangle
\nonumber\\&&\hspace{-0.5cm}
+L_{10} \langle U^\dagger  F^R_{\mu\nu} U F^{L\mu\nu} \rangle
+H_1 \langle F^R_{\mu\nu} F^{R\mu\nu} + F^L_{\mu\nu} F^{L\mu\nu} \rangle
+H_2 \langle \chi^\dagger \chi \rangle\,.
\ea
The matrix $U \in SU(3)$ contains the pseudo-scalars and 
its exponential representation is    
\be
U(\phi) = \exp(i \sqrt{2} \phi/F_0)\,,
\ee
where
\ba
\phi (x) 
 = \, \left( \begin{array}{ccc}
\displaystyle\frac{ \pi_3}{ \sqrt 2} \, + \, \frac{ \eta_8}{ \sqrt 6}
 & \pi^+ & K^+ \\
\pi^- &\displaystyle - \frac{\pi_3}{\sqrt 2} \, + \, \frac{ \eta_8}
{\sqrt 6}    & K^0 \\
K^- & \bar K^0 &\displaystyle - \frac{ 2 \, \eta_8}{\sqrt 6}
\end{array}  \right) .
\ea
The external fields are defined through the covariant derivatives
and field strength tensor as
\be
D_\mu U = \partial_\mu U -i r_\mu U + i U l_\mu \,, \quad
F_{\mu\nu}^{L} = \partial_\mu l_\nu -\partial_\nu l_\mu
-i \left[ l_\mu , l_\nu \right]\,,
\ee
The right-handed and left-handed external fields are 
denoted by  $r_\mu$ and $l_\mu$ respectively.
The Hermitian $3\times3$ matrix $\chi$ contains the scalar ($s$) 
and pseudo-scalar ($p$) external densities and is given as 
$\chi = 2B_0 \left( s + i p\right)$.
The constants $F_{0}$ and $B_{0}$ are related to the pion 
decay constant and quark condensate respectively.
There are however, 10+2 unknown free parameters in the 
Lagrangian ${\cal L}_{4}$ where these effective constants
contain the effects of heavy degrees of freedom and 
can be determined by invoking experimental data 
as well as by Lattice QCD technique. One of the theoretical 
approach, on the other hand, is the application of the 
resonance chiral perturbation which provides an approximate 
estimate of the low energy constants (LECs).
The extention of the chiral Lagrangian to the 
next-to-next-to-leading order is also accomplished \cite{BCE1}. 
At this order there are a large number of LECs, 90+4.

The external scalar field $s$ contains the quark masses and the
mass terms in the lowest order Lagrangian ${\cal L}_2$ can be diagonalized
exactly. In the presence of $m_u\ne m_d$ the physical $\pi^0$ and $\eta$
differ from the triplet and octet states via a lowest-order
mixing angle $\epsilon$
as
\ba
\label{depefpsilon}
\pi_{3} &=& \pi^0 \cos(\epsilon) - \eta  \sin(\epsilon)   
\nonumber\\
\eta_{8} &=& \pi^0 \sin(\epsilon) + \eta \cos(\epsilon) 
\ea
The lowest order mixing angle is 
\ba
\tan(2\epsilon) &=& \frac{\sqrt{3}}{2}\frac{m_d-m_u}{m_s-\hat m}\,,
\nonumber\\ 
\hat m &=& (m_u+m_d)/2\,.
\ea

A review on ChPT to order $p^6$ is \cite{reviewp6}. References to other recent
reviews and lectures can be found there.

\section{Matrix-elements in the presence of mixing}
\label{sect:mixing}

\begin{figure}\begin{center}
\includegraphics[width=0.3\textwidth,clip]{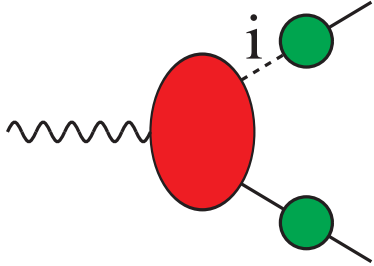}
\end{center}\caption{
The full three-point Green function is represented. 
The oval stands for the amputated three-point Green function and
circles indicate the full two-point functions. 
The solid lines are external mesons 
and the dashed line labeled by i , indicates the sum
over states implied in the external leg where mixing occurs.
The wiggly line indicates the external vector current.}
\label{figgreen_function2}
\end{figure}

For this work we need to work out the matrix elements defined earlier
in the presence of mixing.
These matrix elements can be determined
from three-point Green functions. 
Two of the external legs 
are the meson propagators and the third one is the 
external field. The matrix element is obtained from the Green
function using the  Lehmann-Symanzik-Zimmermann (LSZ) reduction 
formula. The matrix element is 
related to the residue of the Green function in momentum 
space where all the propagators are continued to 
the on-shell mass. 
The case of two point-function with one leg undergoing mixing is 
worked out in \cite{ABT4} and we generalized this 
to a four point-function with mixing on two external legs in \cite{BK}.
In this article we study the form-factors in
$K^{i} \to \pi^{j}$ transitions where, in case of the neutral pion in the 
decay product, mixing should also taken into account. 
Fig.~\ref{figgreen_function2} depicts the three-point Green function 
relevant for this work where we have only considered mixing in one 
external propagator.           
Amplitudes are obtained via
\be
\label{defLSZ}
{\cal A}_{i_1\ldots i_n}
= 
\left( \frac{(-i)^{n}}{\sqrt{Z_{i_1} \ldots Z_{i_n}}} \right)\, 
\prod_{i=1}^n \, 
\lim_{k_i^2\to m_i^2}(k_i^2-m_i^2)\,
G_{i_1\ldots i_{n}}
(k_1,\ldots,k_n)\,.
\ee
This formula shows the general case with n-outgoing particles. 
The function $G_{i_1\ldots i_{n}}$ is the full n-point Green function
where we now express it in terms of the amputated Green functions 
and only two meson propagators to suite for the current article
as follows
\be
\label{decomposition}
G_{43,extv}  =  G_{44}(p^2\approx m_{4\,phys}^2) G_{3i}(p^2\approx m_{3\,phys}^2) {\cal G}_{4i,extv}\,.
\ee
Summation over index i runs over two possibilities of being 
a neutral pion or eta.
4 and 3 are the indices referring to the Kaon and the neutral pion respectively.
${\cal G}_{4i}$ is the amputated Green function that contains both on-shell 
and off-shell Feynman diagrams. The two-point functions are expanded near
the physical poles as
\be
G_{ii}(p^2\approx m_{i\,phys}^2) = \frac{i Z_i}{p^2-m_{i\,phys}^2}\,.
\ee 
The function $Z_{i}$ is called the wavefunction renormalization factor.
The expansion of the off-diagonal two-point functions around the physical
poles is somewhat more involved but can be done in terms of the one-particle
irreducible two-point functions $\Pi_{ij}(m^2)$ and the mass differences
needed in the propagator of the $i$-leg in Fig.~\ref{figgreen_function2}
as explained in \cite{ABT4}. 
We now expand all quantities to the required chiral order and use the fact
that we have exactly diagonalized the lowest Lagrangian to obtain for
the full amplitudes to order $p^6$: 
\ba
\label{fullamplitude}
{\cal A}_{43,extv} &=& {\cal A}_{43,extv}^{(2)}+
      {\cal A}_{43,extv}^{(4)}+{\cal A}_{43,extv}^{(6)}+\cdots\,,
\\
\label{fullp2}
{\cal A}_{43,extv}^{(2)} &=& {\cal G}_{43,extv}^{(2)} \,,
\\
\label{fullp4}
{\cal A}_{43,extv}^{(4)} &=&
{\cal G}_{43,extv}^{(4)} 
- \left( \frac{1}{2} Z_{44}^{(4)} + \frac{1}{2} Z_{33}^{(4)} \right){\cal G}_{43,extv}^{(2)}
-\frac{\Pi_{38}^{(4)}(3)}{\Delta m_2^2} {\cal G}_{48,extv}^{(2)} \,,
\\
\label{fullp6}
{\cal A}_{43,extv}^{(6)} &=&
{\cal G}_{43,extv}^{(6)}
-\frac{1}{2} \left(Z_{33}^{(6)}+ Z_{44}^{(6)} \right)
        {\cal G}_{43,ext}^{(2)}
-\frac{1}{2} \left(Z_{33}^{(4)}+ Z_{44}^{(4)} \right)
      {\cal G}_{43,extv}^{(4)}
\nonumber\\&&
+\frac{3}{8} \left( \left( Z_{33}^{(4)}\right)^2 +\left( Z_{44}^{(4)}\right)^2 
      \right) {\cal G}_{43,ext}^{(2)} 
+ \frac{1}{4} \left( Z_{33}^{(4)}\, Z_{44}^{(4)} \right) {\cal G}_{43,extv}^{(2)}
\nonumber\\&&
+\frac{\Pi_{38}(3)^{(4)}}{\Delta m_2^2}{\cal G}_{48,extv}^{(4)}
+\frac{\Pi_{38}(3)^{(6)}}{\Delta m_2^2}{\cal G}_{48,extv}^{(2)}
+\frac{\Pi_{38}(3)^{(4)}\,\Pi_{88}(3)^{(4)}}{\Delta m_2^2}{\cal G}_{48,extv}^{(2)}
\nonumber\\&&
-\frac{1}{2} \left( Z_{38}^{(4)} \frac{\Pi_{38}(3)^{(4)}}{\Delta m_2^2} \right) {\cal G}_{43,extv}^{(2)}
-\frac{1}{2} \left(  Z_{33}^{(4)}\frac{\Pi_{38}(3)^{(4)}}{\Delta m_2^2}
                    +Z_{44}^{(4)}\frac{\Pi_{38}(3)^{(4)}}{\Delta m_2^2}
              \right)  {\cal G}_{48,extv}^{(2)} 
\,.
\nonumber\\
\ea
The $Z$ and $\Pi$ factors have been valuated earlier \cite{ABT4,BK}
and we thus need to evaluate the various ${\cal G}$ amputated amplitudes.

\section{Analytical results and ratios of form-factors}
\label{sect:ratios}

\begin{figure}\begin{center}
\includegraphics[width=12cm]{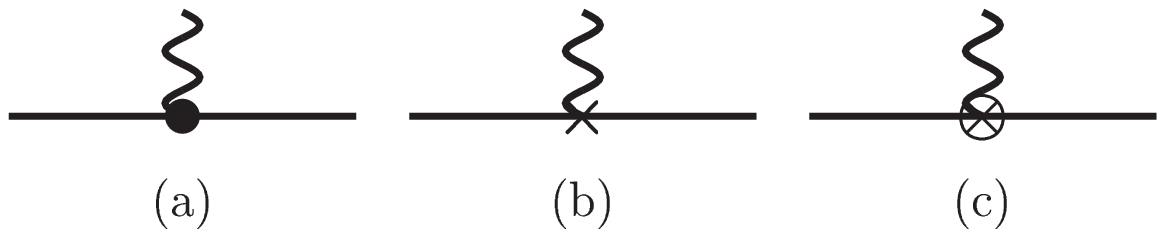}
\end{center}\caption{\label{figtree} The tree level Feynman diagrams for the Kaon
transition form-factors. The wiggly line indicates the insertion of
the vector current, a dot an order $p^2$ vertex, a cross an order $p^4$ vertex
and a crossed circle an order $p^6$ vertex.
}
\end{figure}

\begin{figure}\begin{center}
\includegraphics[width=15cm]{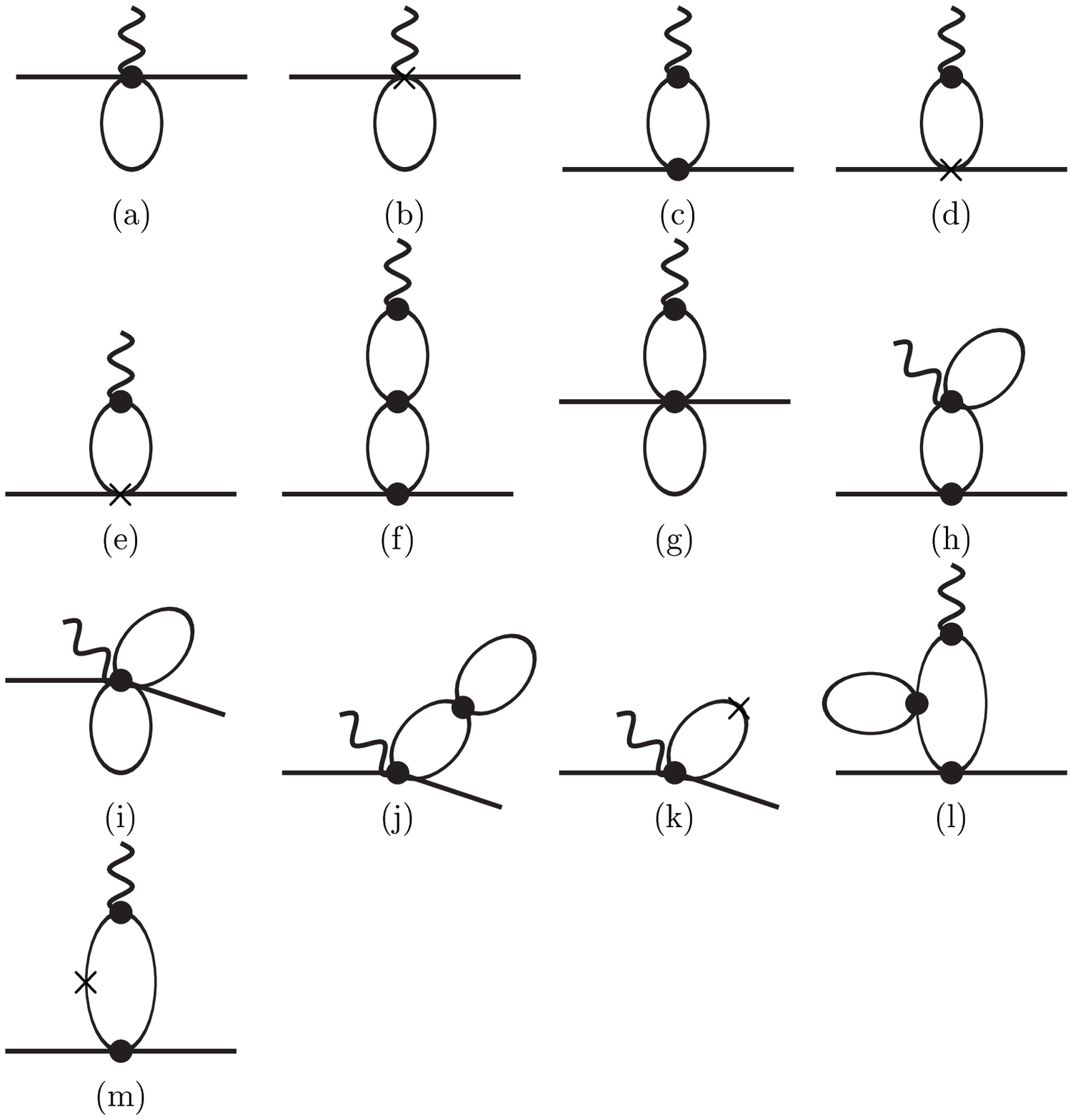}
\end{center}\caption{\label{figoneloop} The one- and two-loop Feynman diagrams for the Kaon
transition form-factors without overlapping divergences.
Notation as in Fig.~\protect\ref{figtree}.}
\end{figure}

\begin{figure}\begin{center}
\includegraphics[width=12cm]{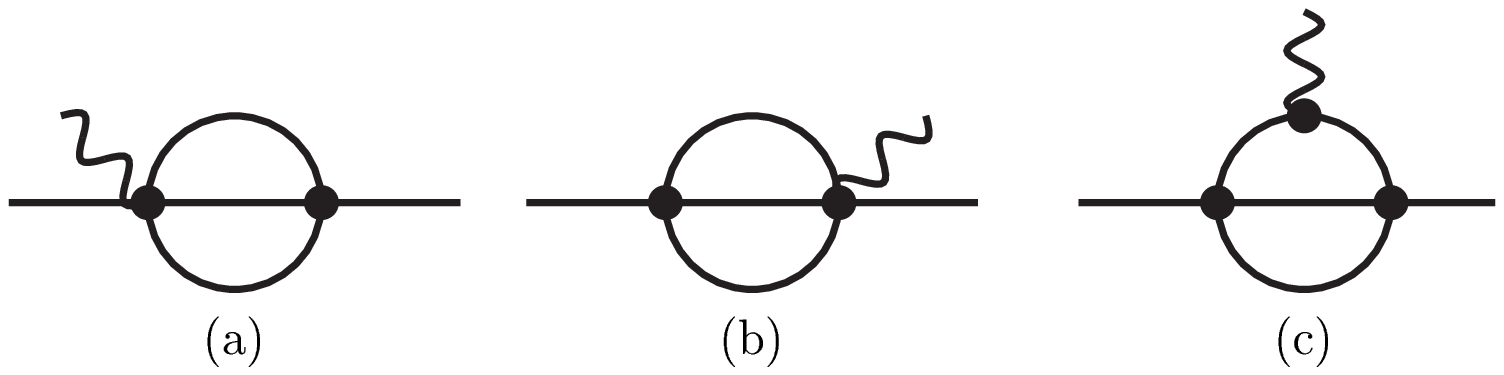}
\end{center}\caption{\label{figtwoloop} The two-loop Feynman diagrams for the Kaon
transition form-factors with overlapping divergences.
Notation as in Fig.~\protect\ref{figtree}.}
\end{figure}

To do the calculation, we need to calculate the tree level diagrams
of Fig.\ref{figtree}, the one- and two-loop diagrams of Fig.~\ref{figoneloop}
and the two-loop diagrams with overlapping divergences of Fig.~\ref{figtwoloop}
with isospin breaking kept in the masses and vertices. These amplitudes should
then be put together with the wave-function renormalization and mixing effects
given in (\ref{fullamplitude}). 
The lowest order expressions are
quite simple. The form-factors $f_-^{K^i\pi^j}$ all vanish
and the others are
\ba
\label{resultLO}
f_+^{K^+\pi^0}(t) &=& \cos\epsilon+\sqrt{3}\sin\epsilon\,,
\nonumber\\
f_+^{K^0\pi^-}(t) &=& 1\,,
\nonumber\\
f_+^{K^+\pi^+}(t) &=& 1\,,
\nonumber\\
f_+^{K^0\pi^0}(t) &=& \cos\epsilon-\sqrt{3}\sin\epsilon\,.
\ea
The NLO expressions agree with the isospin breaking ones calculated
in \cite{GL2,Mescia1} for the $f_+$ form-factors. The isospin breaking in
the $f_-$ and $f_0$ form-factors is new. The NLO results are given
in App.~\ref{App:NLO}. The full NNLO results are very lengthy but
we have performed two independent calculations that are in agreement. All
eight form-factors are also finite using the general subtractions
calculated in \cite{BCE2}. The nonlocal divergences and the
other quantities that can be removed using $\overline{MS}$ subtraction also
cancel as they should. These consistency checks are described in detail in
\cite{BCEGS2}. The loop integrals are computed using the methods
described in \cite{ABT1,BT2}.

The main existing previous work is for $K_{\ell3}$ decays.
Isospin breaking to order $p^4$ for $f_+$ was done
in \cite{GL2} and the electromagnetic parts worked out
in \cite{Cirigliano0}  to order $e^2p^2$.

While this work was in progress, an analysis of the isospin breaking
in the rare decay form-factors $f_+^{K^+\pi^+}$ and $f_+^{K^0\pi^0}$ to NLO
and order $e^2p^2$ appeared \cite{Mescia1}. They also noted that the relation
(\ref{reliso2}) was satisfied but do not seem to have realized it is an
immediate consequence of isospin.

Isospin breaking in $f_-^{K^i\pi^j}$ has not been discussed earlier
within the context of ChPT.

In \cite{GL2} another relation valid to NLO and first order in isospin breaking
was found.
The ratio of form-factors
\be
\label{rel3}
\frac{f_+^{K^0\pi^-}(t)}{f_+^{K^+\pi^0}(t)} = 
\frac{f_+^{K^0\pi^0}(t)}{f_+^{K^+\pi^+}(t)}
\ee
is independent of momenta and can be cleanly predicted in terms of pseudo-scalar
meson masses. The equality follows from the use of (\ref{reliso2}).

Our results satisfy the relations (\ref{reliso1}) and (\ref{reliso2}),
we had to use a large number of relations between the various
integrals to check this and obtained in this way another nontrivial
check on our results. The NLO relation found by \cite{GL2} is no longer true
at NNLO. There are $t$-dependent corrections at order NNLO of
all\footnote{That it was not valid for the $C_i^r$ contributions at 
order $p^6$ was also noticed in \cite{Mescia1}.} types, pure two-loop,
$L_i^r$-dependent and $C_i^r$-dependent ones.
The relation (\ref{rel3}) is also not true for the scalar form-factors
$f_0^{K^i\pi^j}(t)$
nor for $f_-^{K^i\pi^j}(t)$ already at NLO.

We define here also two more ratios for later use, first the ratio of the two
weak decay form-factors
\be
\label{rel4}
r_{0-}(t) = \frac{f_+^{K^+\pi^0}(t)}{f_+^{K^0\pi^-}(t)}
\ee
and second the ratio of the rare to the weak decay with charged pions
in the final state
\be
\label{rel5}
r_K(t) = \frac{f_+^{K^+\pi^+}(t)}{f_+^{K^0\pi^-}(t)}\,.
\ee
We define similarly definitions of $r^0(t)$, $r_{0-}^0(t)$ and $r_K^0(t)$
for ratios of the scalar form-factors $f_0^{K^i\pi^j}(t)$.

\section{Resonance estimate of the contribution from the $C_i^r$}
\label{sect:resonance}

This contribution is the most difficult to estimate. In the isospin limit,
$f_+^{K\pi}(0)$ only depends on the combination
$\left(C_{12}^r+C_{34}^r\right)\left(m_K^2-m_\pi^2\right)^2$ \cite{BT3}
and its estimate is the main uncertainty in the chiral prediction for
$f_+^{K\pi}(0)$. A review can be found in \cite{Kaon07Cirigliano}.
The under lying reason for the factor $\left(m_K^2-m_\pi^2\right)^2$ is
the Ademollo-Gatto theorem\cite{Ademollo}. The reasoning used there
remains valid also in the case with isospin breaking for the form-factors
that do not involve $\pi^0$-$\eta$ mixing. The isospin conserving case
is proportional to $\left(m_s-\hat m\right)^2$, but
including isospin breaking, the form-factor for the
neutral weak decay is proportional to $\left(m_s-m_u\right)^2$ and for
the charged rare decay it is proportional to $\left(m_s-m_d\right)^2$.
The full order $p^6$ tree level contribution in these cases is again
proportional to $C_{12}^r+C_{34}^r-\left(L_5^r\right)^2$ just as 
was found for the isospin
conserved case in \cite{BT3,Cirigliano2}.

The general method we use to estimate the $C_i^r$ is of saturation by a finite
number of resonances introduced by \cite{EGPR,EGLPR}. We use the
vector Lagrangian in the Proca formulation with
parameters as determined in \cite{BCEGS2,BT3}. The scalar effect
was studied in detail in \cite{Cirigliano2} and more generally
in \cite{Cirigliano1}. Some problems with this procedure are discussed
in \cite{BGLP}.

The vector exchange contribution does not contribute to the values at
$t=0$ for $f_+^{K^i\pi^j}(t)$. It does however contribute strongly away from zero.
The estimate we use here for the $C_i^r$ from vector exchange is described
in \cite{BT3}. In particular, the same estimate is in good agreement with
the estimate of the curvature in the pion electromagnetic form-factor which
leads to an experimental determination of\cite{BT2}
\be
-4\left(C_{88}^r-C_{90}^r\right) = \left(0.22\pm0.02\right)\,10^{-3}
\ee
compared with a prediction of $0.26\,10^{-3}$.
This is the part that estimates the contribution from the $C_i^r$
in Fig.~\ref{figfvpkowp6}. The way we have implemented it here is
via the effect on the $C_i^r$ directly as given in \cite{Kampf}.

Second, we take into account the contribution from
the singlet pseudo-scalar degree of freedom $P_1$.
We use the simple Lagrangian
\be
\label{lageta}
{\cal L}_{\eta^\prime} = \frac{1}{2}\partial_\mu P_1\partial^\mu P_1
-\frac{1}{2}M_{\eta^\prime}^2 P_1^2
+i\tilde d_m P_1\langle\chi_- \rangle\,.
\ee
Integrating out $P_1$ leads to the order $p^4$ term with $L_7$
and the order $p^6$ Lagrangian
\be
\label{etaprime}
{\cal L}_{\eta'} =  
- \frac{\tilde{d}_m^2}{2 M^4_{\eta'}} 
\partial_\mu \langle \chi_- \rangle \partial^\mu \langle \chi_- \rangle
~\mbox{with}~ \tilde{d}_m = 20 \, \, \mbox{MeV}. 
\ee
The latter was rewritten in general in terms of the basis of operators
of \cite{BCE1} in \cite{BK}.
The result is\footnote{This was derived by the authors of
 \cite{ABT3} but not included in the final manuscript.
It also agrees with the expression shown by Kaiser\cite{Kaiser}.}
\ba
\label{relOi}
\partial_\mu\langle\chi_-\rangle\partial_\mu\langle\chi_-\rangle
&=& O_{18}+\frac{2}{9}O_{19}-\frac{1}{3}O_{20}+\frac{1}{3}O_{21}
+ 2 O_{27}+\frac{2}{3}O_{31}-\frac{1}{3}O_{32}+\frac{1}{3}O_{33}
\nonumber\\&&
-2O_{35}
+ O_{37}-\frac{8}{3}O_{94}\,.
\ea
The result is that the singlet $P_1$ contributes via the order $p^6$
constants $C_i^r$ also to the isospin breaking in the values for 
$f_+^{K^i\pi^j}(0)$
but it does so only via $\pi^0$-$\eta$ mixing.
The numerical result is
\ba
\label{resultetap}
\left.f_+^{K^+\pi^0}(0)\right|_{P_1} &=& 0.00065\,,
\nonumber\\
\left.f_+^{K^0\pi^0}(0)\right|_{P_1} &=& -0.00065\,.
\ea

\section{Numerical results}
\label{sect:numerics}

\subsection{Input parameters}
\label{sectinputs}

For the masses we use the particle data book masses except for the eta where
we use for consistency the value 547.3~MeV.
The input values for the order $p^4$ constants $L_i^r$ we use are fit 10
of \cite{ABT4}. This fit used the $K_{e4}$ data from E865, and input values
$m_s/\hat m=24$ and $F_K/F_\pi=1.22$. For the masses it
used the physical masses.
Electromagnetic corrections to the Kaon mass were included with the
estimate of the violation of Dashen's theorem of \cite{BP} included.
An extensive discussion of this fit can be found in \cite{ABT3} using
the older $K_{e4}$ data and working fully in the isospin limit.

We will always quote results for
the isospin conserving formulas of \cite{BT2} where the kaon mass is taken to
be the mass of the kaon involved and similarly, we use for the pion mass
the mass of the particle involved in the matrix element.
For the results with the formulas including isospin breaking, we have used
for the Kaons their physical masses but for both charged and neutral pion the
same mass, since to first order in $m_u-m_d$ these have the same mass.
We have always taken the mass of the final state pion involved
in the matrix element. The reason for this choice is to always have the
kinematics right in the matrix elements. The effect of changing the pion mass
can be judged by looking at the results for the isospin symmetric formulae
which we quote for different input Kaon and pion masses
in Tab.~\ref{tabfp0inv}..
The order $p^6$ constants $C_i^r$ have been put to zero at the
scale $\mu=770$~MeV unless otherwise noted in Sect.~\ref{sect:fp0}.
In Sects. \ref{sect:fpt}, \ref{sect:fot} and \ref{sect:CT}
we have put the $C_i^r$ at the value estimated by vector and singlet
pseudo-scalar exchange at $\mu=700$~MeV.

The main fit 10 with Dashen's violation gave $m_u/m_d = 0.45$ while removing
the violation of Dashen's theorem gave  $m_u/m_d = 0.52$. The standard values
without order $p^6$ and without violation of Dashen's theorem
gave $m_u/m_d=0.585$ \cite{ABT4}. These values, together with the input value
for $m_s/\hat m$ correspond to
$\sin\epsilon = 0.0143,$ 0.0119 and 0.00986 respectively. This can be compared
with the value of $0.0106\pm0.0008$ used in \cite{Mescia1} which used the
input neglecting order $p^6$ effects. Note however that the recent
evaluation from $\eta\to3\pi$ \cite{BK} leads to somewhat different values.

\subsection{$f_+^{K^i\pi^j}(0)$}
\label{sect:fp0}

Here we give the results for the form-factor values at zero. In
Tab.~\ref{tabfp0inv} we first show the results for the isospin conserving
formula of \cite{BT3}. Here the only isospin breaking effect is the different
kaon and pion mass used as described in Sect.~\ref{sectinputs}.
The results for the charged and neutral weak decay
are in agreement with \cite{BT3}. We have in fact checked that the formulas
including isospin breaking numerically agree with the isospin conserving
formula if the masses are set to the same isospin conserving masses
and $\sin\epsilon=0$. As is clear from the numbers in Tab.~\ref{tabfp0inv},
the isospin breaking effects from varying the masses in the loops is quite
small.
\begin{table}\begin{center}
\begin{tabular}{|c|cccc|}
\hline
        & $f_+^{K^+\pi^0}$ & $f_+^{K^0\pi^-}$ & $f_+^{K^+\pi^+}$ & $f_+^{K^0\pi^0}$\\
\hline
  order $p^2$            &   1.00000 &   1.00000 &   1.00000 &    1.00000 \\
  order $p^4$            &$-$0.02276 &$-$0.02266 &$-$0.02226 & $-$0.02316 \\
  order $p^6$            &   0.01423 &   0.01462 &   0.01406 &    0.01480 \\
  $p^6$ 2-loop           &   0.01104 &   0.01130 &   0.01090 &    0.01145 \\
  $p^6$ $L_i^r$-dependent &  0.00320 &   0.00332 &   0.00316 &    0.00336 \\
\hline
sum of $p^2$, $p^4$ and $p^6$& 0.99156 &  0.99196 & 0.99180   & 0.99164\\
\hline
\end{tabular}
\end{center}\caption{\label{tabfp0inv} The different contributions to $f_+^{K^i\pi^j}(0)$
using the isospin \emph{conserving} amplitudes of \protect\cite{BT3}.
We have also shown the break-up of the order $p^6$ expressions in the pure
two-loop part and the $L_i^r$-dependent part. The part depending on the $C_i^r$
is \emph{not} included.}
\end{table}

In contrast, we have shown the equivalent set of values for our
amplitudes including isospin violation. It can be seen that effect is much
larger for the amplitudes with a neutral pion in the final state. That is, as
can already be seen at lowest order, pion-eta mixing is important for this
decay. The values in Tab.~\ref{tabfp0} are with $m_u/m_d=0.45$ or
$\sin\epsilon=0.01429$. 
\begin{table}\begin{center}
\begin{tabular}{|c|cccc|}
\hline
        & $f_+^{K^+\pi^0}$ & $f_+^{K^0\pi^-}$ & $f_+^{K^+\pi^+}$ & $f_+^{K^0\pi^0}$\\
\hline
  order $p^2$            &   1.02465 &   1.00000 &   1.00000 &   0.97514\\
  order $p^4$            &$-$0.01775 &$-$0.02292 &$-$0.02197 &$-$0.02838\\
  order $p^6$            &   0.00809 &   0.01470 &   0.01391 &   0.02095\\
  $p^6$ 2-loop           &   0.00159 &   0.01145 &   0.01081 &   0.02092\\
  $p^6$ $L_i^r$-dependent &  0.00650 &   0.00325 &   0.00309 &   0.00004\\
\hline
sum of $p^2$, $p^4$ and $p^6$&1.01499 &   0.99177 &   0.99194 &   0.96772\\
\hline
\end{tabular}
\end{center}\caption{\label{tabfp0} The different contributions to $f_+^{K^i\pi^j}(0)$
using the amplitudes \emph{including isospin breaking}.
We have also shown the break-up of the order $p^6$ expressions in the pure
two-loop part and the $L_i^r$-dependent part. The part depending on the $C_i^r$
is \emph{not} included. We used here $m_u/m_d=0.45$ corresponding
to the two-loop fit of \protect\cite{ABT4} including Dashen's theorem
violations.}
\end{table}

To show the variation with the input for $m_u/m_d$,
we show in Tab.~\ref{tabfp02} using the same inputs as for Tab.~\ref{tabfp0}
but with $m_u/m_d=0.585$ or $\sin\epsilon=0.009857$. This corresponds to
the fit for $m_u/m_d$ without violations of Dashen's theorem and
using order $p^4$ expressions.
Our results, except for the lowest order in (\ref{resultLO}), are explicitly
linear in $\sin\epsilon$.
\begin{table}\begin{center}
\begin{tabular}{|c|cccc|}
\hline
        & $f_+^{K^+\pi^0}$ & $f_+^{K^0\pi^-}$ & $f_+^{K^+\pi^+}$ & $f_+^{K^0\pi^0}$\\
\hline
  order $p^2$            &   1.01702 &   1.00000 &   1.00000 &   0.98288\\
  order $p^4$            &$-$0.01931 &$-$0.02282 &$-$0.02202 &$-$0.02675\\
  order $p^6$            &   0.00986 &   0.01467 &   0.01395 &   0.01919\\
  $p^6$ 2-loop           &   0.00435 &   0.01142 &   0.01084 &   0.01815\\
  $p^6$ $L_i^r$-dependent &  0.00551 &   0.00325 &   0.00311 &   0.00104\\
\hline
sum of $p^2$, $p^4$ and $p^6$&1.00757 &   0.99186 &   0.99193 &   0.97532\\
\hline
\end{tabular}
\end{center}\caption{\label{tabfp02} The different contributions to $f_+^{K^i\pi^j}(0)$
using the amplitudes \emph{including isospin breaking}.
We have also shown the break-up of the order $p^6$ expressions in the pure
two-loop part and the $L_i^r$-dependent part. The part depending on the $C_i^r$
is \emph{not} included. We used here $m_u/m_d=0.585$ corresponding
to the one-loop fit of \protect\cite{ABT4} without violations of Dashen's
 theorem.}
\end{table}
The  numbers are slightly different from the preliminary results quoted in
\cite{Kaon07}. This is due to a slightly different way of treating the pion
masses.

Using the results of Tab.~\ref{tabfp0} we can also quote numerical results for
the various ratios defined earlier at the point $t=0$.
First the ratio of charged to neutral weak decay.
This is
\be
\label{resultr0-}
r_{0-}(0) = 1.02465+0.00587-0.00711 = 1.02341\,,
\ee
where we see that the order $p^6$ contributions lower the result
and essentially cancel the enhancement from the ratio at order $p^4$.
If we add the contribution from singlet $P_1$ exchange we obtain 
$r_{0-}=1.024068$.
However, compared to the old order $p^4$
value, we get again an enhancement due to the
larger value of $\sin\epsilon$ obtained from the order $p^6$ fit.
We showed the contributions to the ratio at order $p^2$, $p^4$ and $p^6$.
This should be compared to the experimental ratio as determined
from the global FLAVIAnet fit \cite{Kaon07flavia}
\be
r_{0-exp} = 1+\Delta_{SU(2)} = 1.0284\pm0.0040\,.
\ee
As we see, we obtain a reasonable agreement.

We can also look at the double ratio $r$ from (\ref{reliso2}). Our formulas
satisfy it exactly. The main numerical source of the difference
at higher orders
results from the fact that we used a different pion mass in the denominator
and the numerator. The result is
\be
r = 0.99918-0.00161+0.00085 = 0.99842\,,
\ee
where a fairly sizable cancellation happens between the order $p^4$
and order $p^6$ contributions. 
We again showed the contributions to the ratio at order $p^2$, $p^4$ and $p^6$.

The final ratio, of weak to rare decays with a charged pion in the final state
is
\be
r_K = 1.00000+0.00097-0.00080 = 1.00017\,.
\ee
The three numbers in the middle part are
once more the contributions to the ratio at order $p^2$, $p^4$ and $p^6$.
Once more, we see a significant cancellation between the order $p^4$ and
$p^6$ contributions.

\subsection{$f_+^{K^i\pi^j}(t)$}
\label{sect:fpt}

\begin{figure}\begin{center}
\includegraphics[width=12cm]{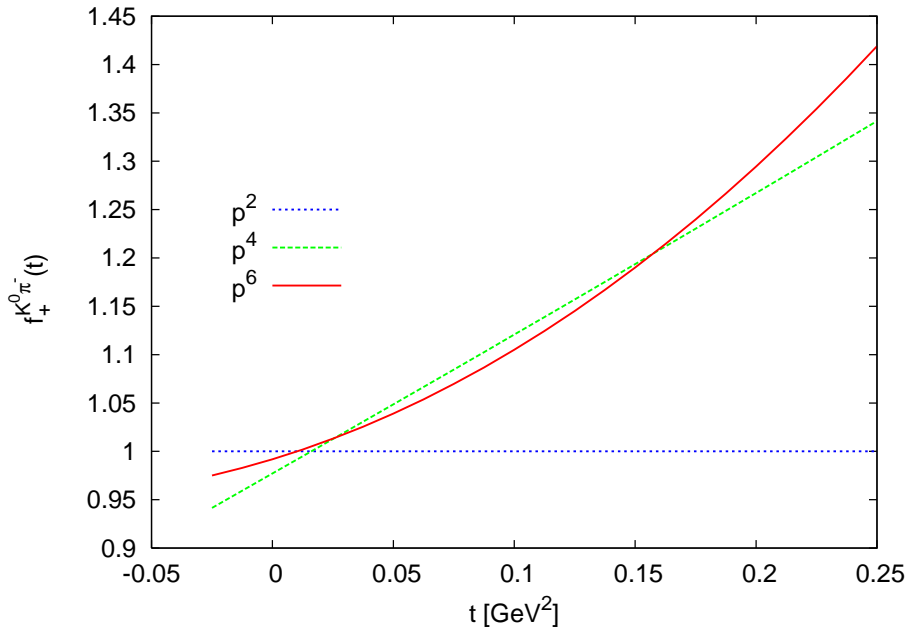}
\end{center}\caption{\label{figfvpkow} The form-factor $f_+^{K^0\pi^-}(t)$ 
as a function
of $t$. Shown are the lowest order ($p^2$), NLO ($p^4$) and NNLO result ($p^6$).
Isospin breaking is included.
}
\end{figure}

\begin{figure}\begin{center}
\includegraphics[width=12cm]{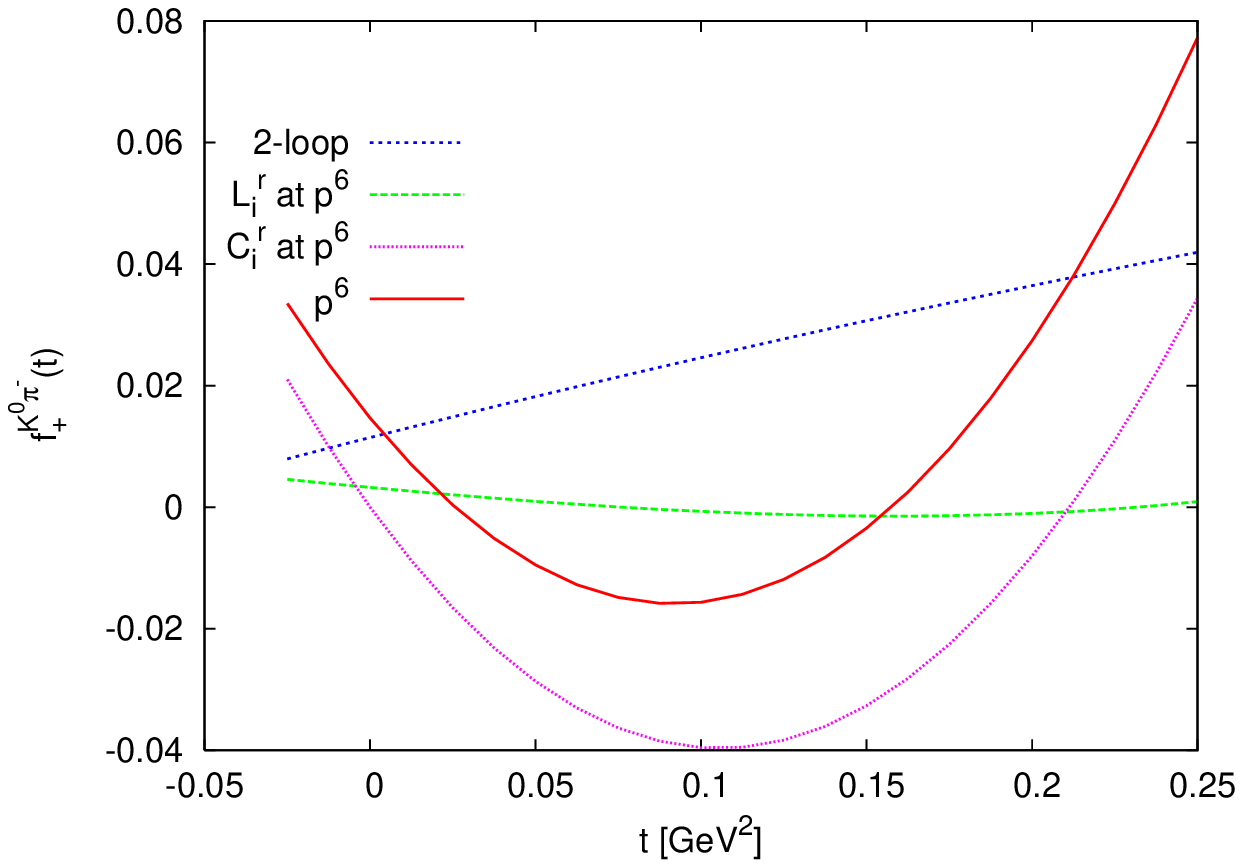}
\end{center}\caption{\label{figfvpkowp6} The form-factor $f_+^{K^0\pi^-}(t)$
 as a function
of $t$. Shown are the full order $p^6$ contribution and its three
constituent parts, the pure two-loop contribution, the
$L_i^r$-dependent part and the $C_i^r$-dependent part.
The contribution to the quadratic slope comes mainly from the $C_i^r$ dependent
part but that is fixed from the pion electromagnetic form-factor
\protect\cite{BT2}. Isospin breaking is included.}
\end{figure}

In this subsection we show the results as a function of $t$ for the
the $f_+^{K^i\pi^j}$ form-factors. We first show the case for the neutral weak
decay in Figs.~\ref{figfvpkow} and \ref{figfvpkowp6}. Fig.~\ref{figfvpkow}
shows the result to lowest order, NLO and NNLO. It can be seen that there is
a nice convergence in the entire region shown.

We show the various subparts of the order $p^6$ contribution in
Fig.~\ref{figfvpkowp6}. The contributions shown are the two-loop
contribution, the part dependent on the order $p^4$ LECs $L_i^r$ as well as
the part that depends on the order $p^6$ LECs $C_i^r$.

\begin{figure}\begin{center}
\includegraphics[width=12cm]{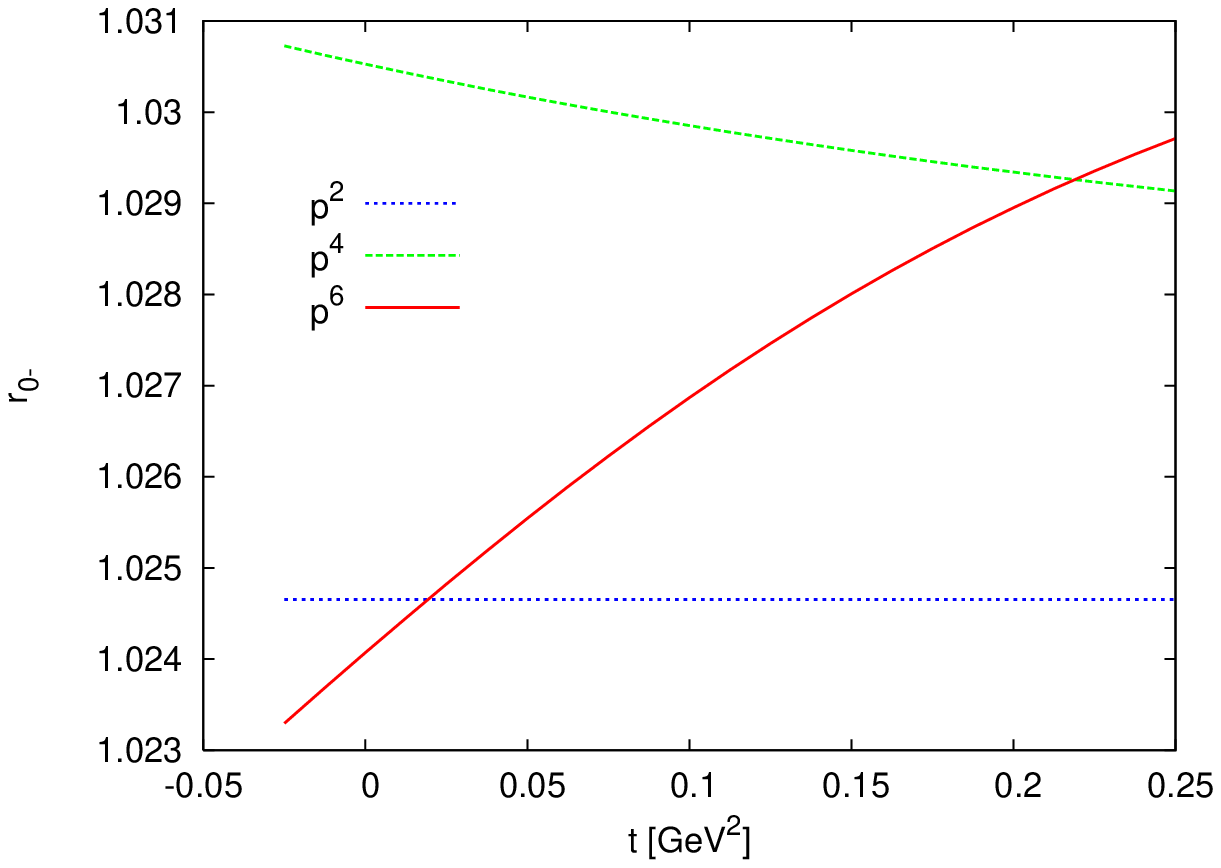}
\end{center}\caption{\label{figr0+} The ratio $r_{0-}$ as defined in (\protect\ref{rel4})
as a function of $t$. This is the ratio of the charged to neutral weak decay.
The $t$ dependence for the NLO result is higher order in
isospin breaking but is first order at NNLO.}
\end{figure}

\begin{figure}\begin{center}
\includegraphics[width=12cm]{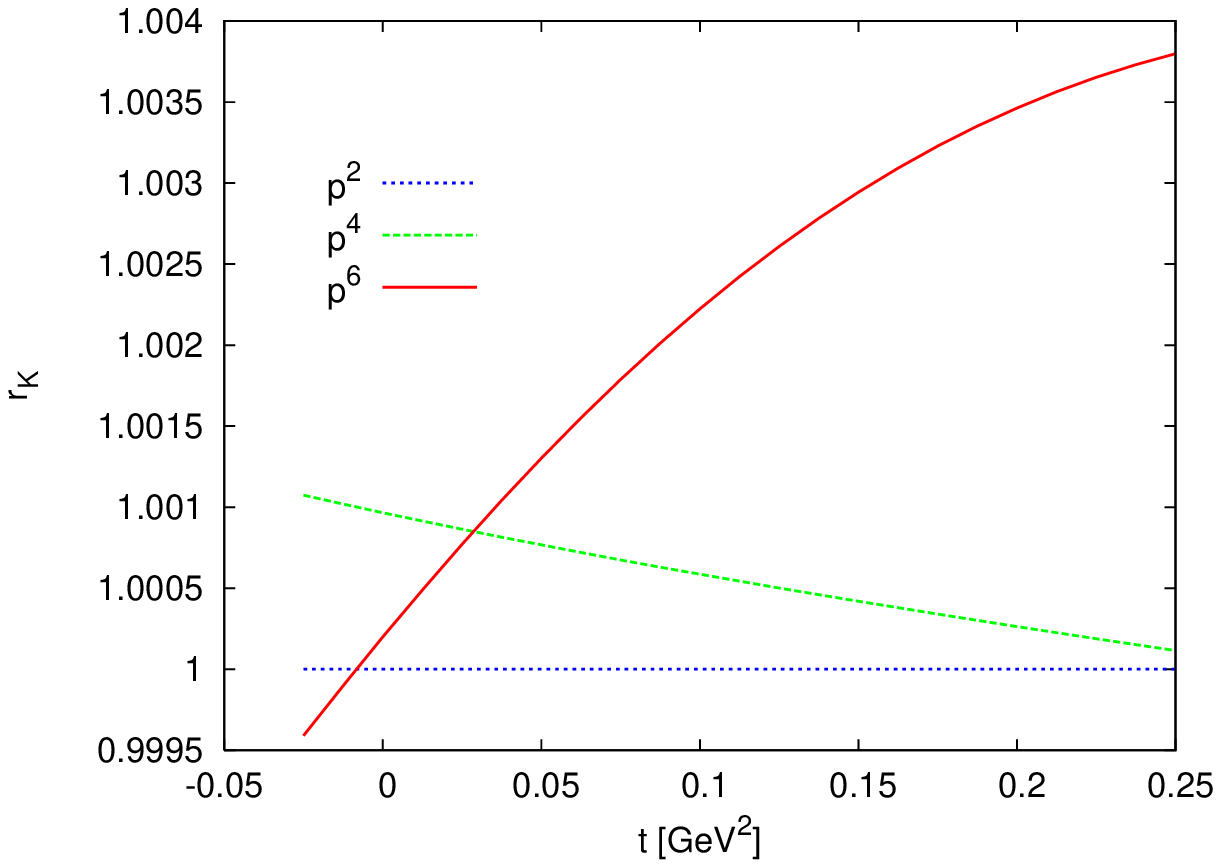}
\end{center}\caption{\label{figrK} The ratio $r_K$ as defined in (\protect\ref{rel5})
as a function of $t$. This is the ratio of the charged rare to neutral weak
decay.}
\end{figure}

The results shown so far for $f_+^{K^0\pi^-}$
are essentially the same as those in the isospin limit
of \cite{BT3}. We have included isospin breaking but it is a rather small
effect for this form-factor. Rather than showing similar plots for the other
three form-factors we show here the ratios as a function of $t$. First we show
the variation of the full ratio $r$ as a function of $t$. The ratio $r$
is somewhat more different from one
 than naively expected since we included different pion masses.
\begin{figure}\begin{center}
\includegraphics[width=12cm]{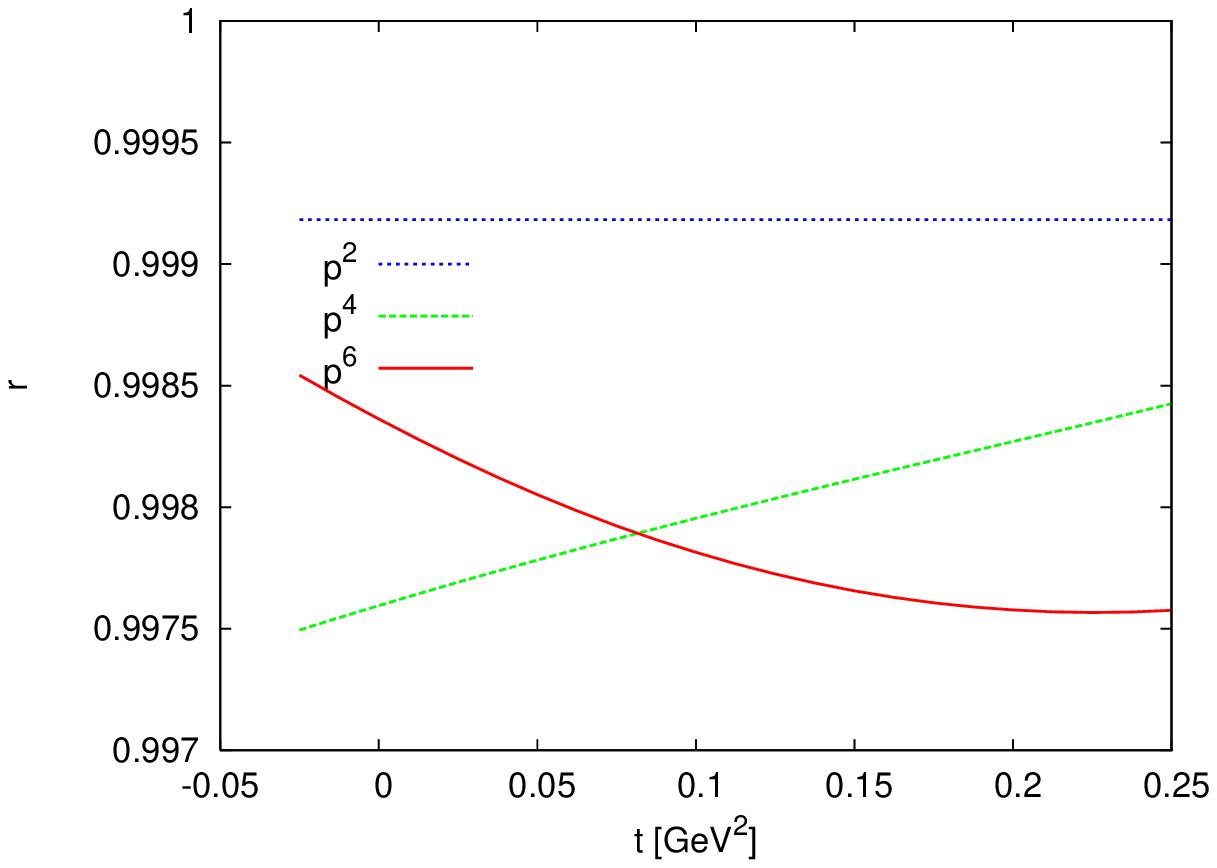}
\end{center}\caption{\label{figr} The ratio $r$ as defined in (\protect\ref{reliso2})
as a function of $t$. Both the deviation from 1 and the $t$ dependence
are effects of higher order in isospin breaking.}
\end{figure}
The ratio $r_{0-}$ defined in (\ref{rel4}) is shown as a function of $t$
in Fig.~\ref{figr0+}. It was shown in \cite{GL2} that at NLO this
ratio is independent of $t$. We have checked that this is no longer true
at NNLO as is clearly visible in the figure. However, there is clearly
no sign of an anomalously large isospin breaking effect in this ratio. 
We also show the similar ratio for the charged rare to neutral weak decay,
$r_K$ as defined in (\ref{rel5}) in Fig.~\ref{figrK}.

\subsection{$f_0^{K^i\pi^j}(t)$}
\label{sect:fot}

\begin{figure}\begin{center}
\includegraphics[width=12cm]{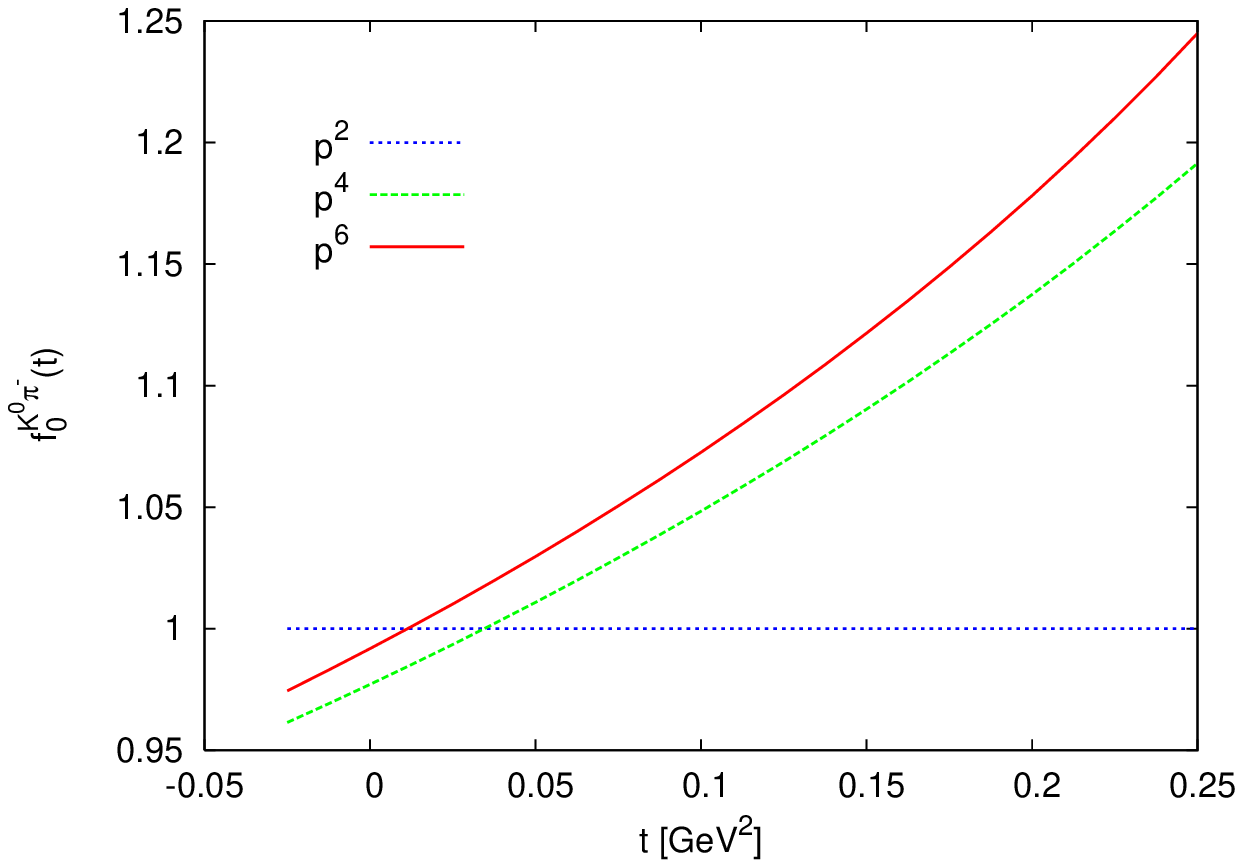}
\end{center}\caption{\label{figfvokow} The form-factor $f_0^{K^0\pi^-}(t)$ as a function
of $t$. Shown are the lowest order ($p^2$), NLO ($p^4$) and NNLO result ($p^6$).
Isospin breaking is included.
}
\end{figure}

\begin{figure}\begin{center}
\includegraphics[width=12cm]{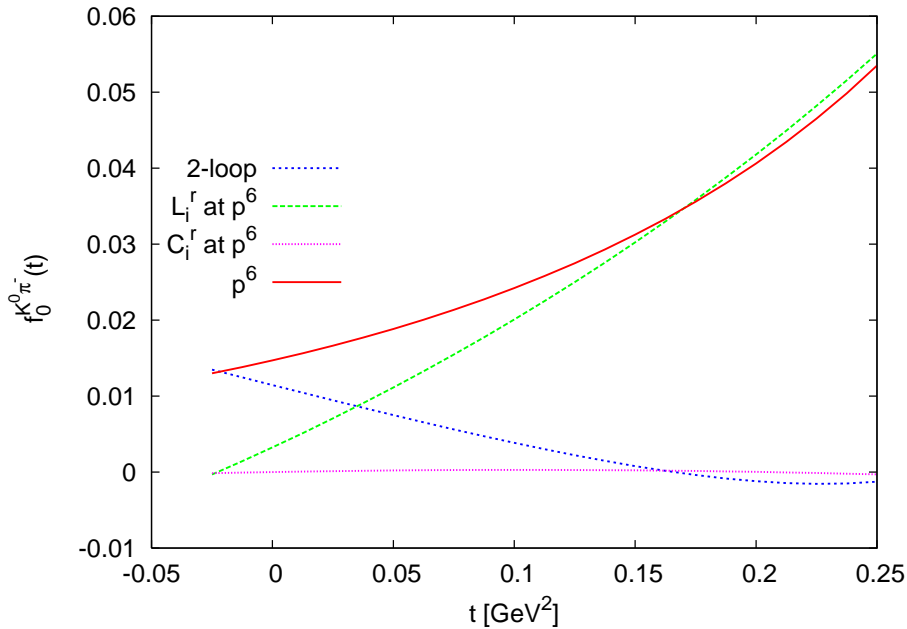}
\end{center}\caption{\label{figfvokowp6} The form-factor $f_0^{K^0\pi^-}(t)$ as a function
of $t$. Shown are the full order $p^6$ contribution and its three
constituent parts, the pure two-loop contribution, the
$L_i^r$-dependent part and the $C_i^r$-dependent part.
Isospin breaking is included.}
\end{figure}

In this subsection we show the results as a function of $t$ for the
the $f_0^{K^i\pi^j}$ form-factors. We first show the case for the neutral weak
decay in Figs.~\ref{figfvokow} and \ref{figfvokowp6}. Fig.~\ref{figfvokow}
shows the result to lowest order, NLO and NNLO. It can be seen that there is
a nice convergence in the entire region shown.

We show the various subparts of the order $p^6$ contribution in
Fig.~\ref{figfvokowp6}. The contributions shown are the two-loop
contribution, the part dependent on the order $p^4$ LECs $L_i^r$ as well as
the part that depends on the order $p^6$ LECs $C_i^r$. The latter is essentially
zero here since the vector exchange contribution to the scalar form-factor
vanishes to the order considered here and the singlet pseudo-scalar doesn't
contribute either. A scalar exchange would contribute but we have not included
such an estimate here. The curvature visible is here mainly coming from
the loops.

\begin{figure}\begin{center}
\includegraphics[width=12cm]{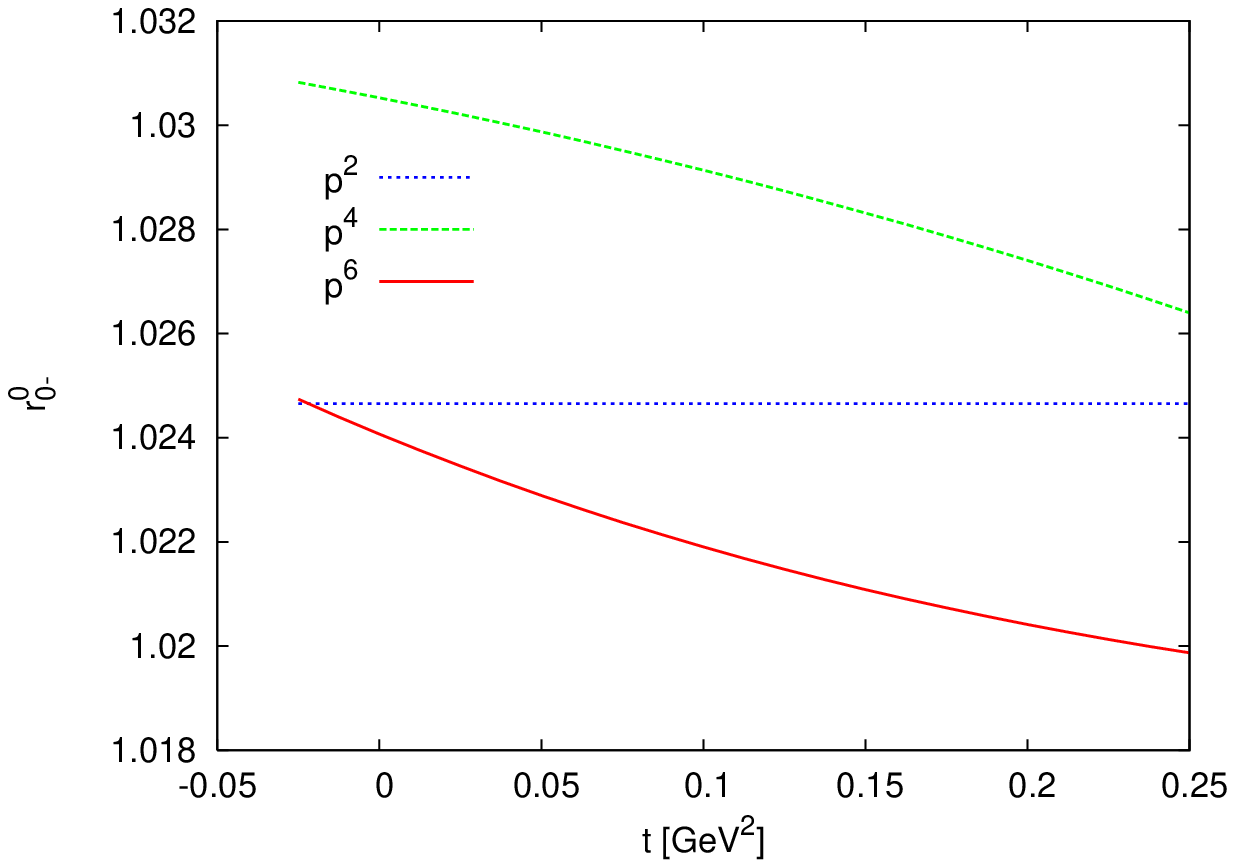}
\end{center}\caption{\label{figr0+o} The ratio $r_{0-}^0$ as defined in (\protect\ref{rel4})
but for the scalar form-factor
as a function of $t$. This is the ratio of the charged to neutral weak decay.
The $t$ dependence is first order in isospin breaking both at NLO and NNLO.}
\end{figure}

\begin{figure}\begin{center}
\includegraphics[width=12cm]{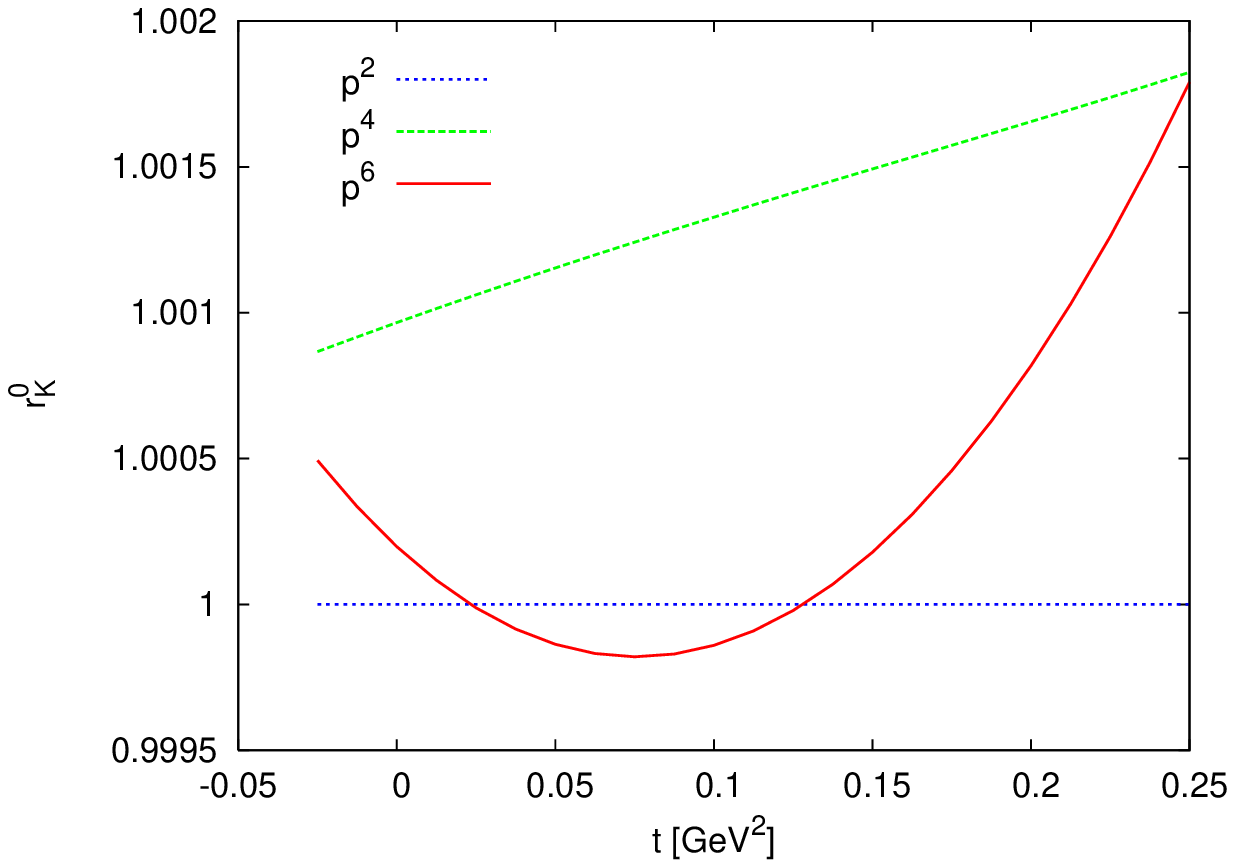}
\end{center}\caption{\label{figrKo} The ratio $r_K^0$ as defined in 
(\protect\ref{rel5})
but for the scalar form-factor
as a function of $t$. This is the ratio of the charged rare to neutral weak
decay.}
\end{figure}

The results shown so far for $f_0^{K^0\pi^-}$
are essentially the same as those in the isospin limit
of \cite{BT3}. We have included isospin breaking but it is a rather small
effect for this form-factor. Rather than showing similar plots for the other
three form-factors we show here the ratios as a function of $t$. First we show
the variation of the full ratio $r^0$ as a function of $t$. The ratio $r^0$
is somewhat
larger than naively expected since we included different pion masses
\begin{figure}\begin{center}
\includegraphics[width=12cm]{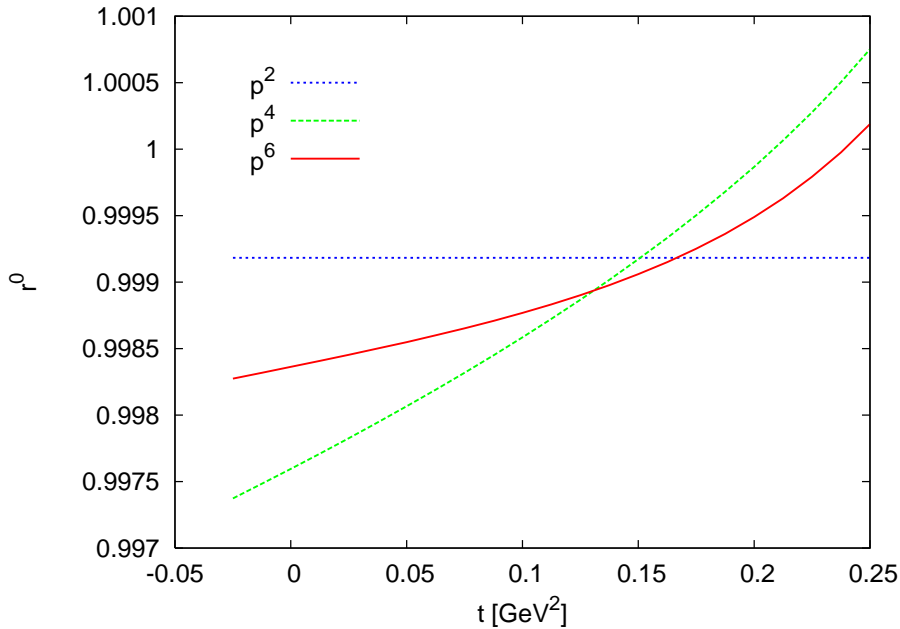}
\end{center}\caption{\label{figro} The ratio $r^0$ as defined in (\protect\ref{reliso2})
bit for the scalar form-factor
as a function of $t$. Both the deviation from 1 and the $t$ dependence
are effects of higher order in isospin breaking.}
\end{figure}
The ratio $r_{0-}^0$ defined in (\ref{rel4}) 
but for the scalar form-factor
is shown as a function of $t$
in Fig.~\ref{figr0+o}. This ratio can be $t$-dependent already at NLO which is
clearly visible.
However, there is 
no sign of an anomalously large isospin breaking effect in this ratio. 
We also show the similar ratio for the charged rare to neutral weak decay,
$r_K^0$ as defined in (\ref{rel5}) in Fig.~\ref{figrKo}.
The scalar form-factors are not needed for the weak decays to an electron or
the rare decays to a neutrino-antineutrino pair. They do contribute to the
weak decays to a muon and the rare decays with a muon-anti-muon pair
via a the axial current couplings to the latter from the short-distance
contributions.

\subsection{Callan-Treiman point}
\label{sect:CT}

The Callan-Treiman relation \cite{CT1} states that the scalar form-factor
at $t=m_K^2-m_\pi^2$ satisfies
\be
f_0\left(m_K^2-m_\pi^2\right) = \frac{F_K}{F_\pi}+{\cal O}(m_u,m_d)\,.
\ee
This relation is derived using current algebra in the up-down--sector
and should thus have rather small corrections of order $m_\pi^2$.
The relation is exact when the up and down quark masses are zero.
The correction at NLO was worked out in \cite{GL2} and found to be
\be
\Delta_{CT} \equiv f_0\left(m_K^2-m_\pi^2\right) - \frac{F_K}{F_\pi}
= -3.5\,10^{-3}\cite{GL2}\,.
\ee
The correction in the isospin limit at NNLO was never presented in \cite{BT3}.
We have calculated this and also for the four different amplitudes at order
$p^4$ but for the $C_i^r$ estimates we have used the vector and
pseudo-scalar singlet contributions as described in Sect.~\ref{sect:resonance}.

The inputs we used produce $F_K/F_\pi=1.22$, which is what we have subtracted
from the $f_0^{K^i\pi^j}\left(m_{K^i}^2-m_{\pi^j}^2\right)$ in the numbers quoted
below.
The isospin symmetric expression with $m_K^2=m_{K^0}^2$ and $m_\pi^2=m_{\pi^+}^2$
gives
\be
\Delta_{CT} =  -6.2\,10^{-3}\,.
\ee
As we see there is a substantial NNLO correction. Note that we did not include
the contributions from nonzero $C_{12}^r$, $C_{34}^r$ in this expression.
These read
\be
\left.\Delta_{CT}\right|_{C_i^r}
 = \frac{16}{F_\pi^4}\left(2 C_{12}^r+C_{34}^r\right)
m_\pi^2\left(m_K^2-m_\pi^2\right)\,.
\ee
It is clear that this satisfies the Callan-Treiman theorem.
Notice that it is the same
combination of order $p^6$ LECs that shows up in the scalar slope when
the part via $F_K/F_\pi$ is subtracted as in Eq.~(5.2) of \cite{BT3}. 

For the expression including isospin breaking we simply present the
numerical results directly
\ba
\Delta_{CT}^{K^+\pi^0} &=&  15.1\,10^{-3}\,,
\nonumber\\
\Delta_{CT}^{K^0\pi^-} &=&  -5.6\,10^{-3}\,,
\nonumber\\
\Delta_{CT}^{K^+\pi^+} &=&  -9.4\,10^{-3}\,,
\nonumber\\
\Delta_{CT}^{K^0\pi^0} &=&  -26.4\,10^{-3}\,.
\ea

Recently, Leutwyler discussed the experimental measurements and the
extrapolation to the Callan-Treiman point \cite{Leutwyler}.
The results we obtained are clearly not sufficient to explain the
large value of $\Delta_{CT}=-0.071\pm0.014_{NA48}\pm0.002_{theo}\pm0.005_{ext}$
 observed by NA48 in the charged weak decay\cite{NA48} but
are in reasonable agreement withe one observed by KLOE for the neutral weak
decay \cite{KLOE}.

\section{Conclusions}
\label{sect:conclusions}

In this paper we have calculated all the vector form-factors of Kaon to Pion
transitions to first order in the quark mass difference $m_u-m_d$
and to NNLO in ChPT. We have thus calculated the eight different form-factors
defined in Eqs.~(\ref{def+0}-\ref{def00}) to order $p^6(m_u-m_d)$.
This complements the earlier calculations to order $p^4(m_u-m_d)$
done for the $f_+$ form-factors in \cite{GL2} and \cite{Mescia1}
and to order $p^6$ in the isospin limit for the vector and scalar
form-factor.

What we find in all cases is the the NNLO results diminish the effects of
isospin breaking but due to the change in $m_u/m_d$ from a NLO to a NNLO
fit the total effect is to increase the isospin breaking in the form-factors.
This goes some way towards reconciling the determinations of $V_{us}$ from
the charged and neutral weak $K_{\ell3}$ decays but does not explain the
full difference. We have also calculated isospin breaking in all the scalar
form-factors. Here again, the effects are sizable but not unexpectedly
large. In particular, they are not large enough to explain the discrepancy
with the Callan-Treiman point observed by NA48\cite{NA48}.

We have presented numerical results for the values at $t=0$
and for the $t$-dependence as well as for various ratios of the form-factors.
In particular, we have shown that the relations (\ref{reliso1}) and
(\ref{reliso2}) are valid to all orders in ChPT and to first order in $m_u-m_d$.
We presented numerical results for the ratios $r$, $r_{0-}$ and $r_K$
as well as for their equivalents for the scalar form-factors.

\section*{Acknowledgments}

This work is supported in part by the European Commission RTN network,
Contract MRTN-CT-2006-035482  (FLAVIAnet), 
the European Community-Research Infrastructure
Activity Contract RII3-CT-2004-506078 (HadronPhysics) and
the Swedish Research Council.

\appendix

\section{The order $p^4$ expression}
\label{App:NLO}

In this appendix we explicitly write out our order $p^4$ results. 
We have checked that they agree with the published results for the $f_+$
form-factors of \cite{GL2} and \cite{Mescia1}.
They also satsify the relation (\ref{reliso2}) when the integrals are expanded
to obtain a common Kaon mass,
 but we have quoted all
eight formfactors here since by rewriting one can move things between the 
order $p^4$ and $p^6$. The expressions quoted here are the ones we used
to define the order $p^4$ part. The integrals used below are
the standard one-loop integrals defined in many places, see e.g. \cite{ABT1}.
\ba
f_+^{K^+\pi^0(4)}(t) &=&
%\nonumber\\&&
        \frac{1}{F_\pi^2\left(m_{\pi^0}^2-m_{\eta}^2\right)} \, \Big(
%\nonumber\\&&
          - 2/3\,\overline{A}(m_{K^+}^2)\,m_{K^+}^2
%\nonumber\\&&
          - 1/3\,\overline{A}(m_{K^+}^2)\,m_{\pi^0}^2
\nonumber\\&&
          + 2/3\,\overline{A}(m_{K^0}^2)\,m_{K^+}^2
%\nonumber\\&&
          + 1/3\,\overline{A}(m_{K^0}^2)\,m_{\pi^0}^2
%\nonumber\\&&
          \Big)
\nonumber\\&&
       + \frac{\sin\epsilon}{\sqrt{3}\,F_\pi^2\,\left(m_{\pi^0}^2-m_{\eta}^2
      \right)} \, \Big(
%\nonumber\\&&
          + 128\,m_{K^+}^4\,L^r_{8}
%\nonumber\\&&
          + 384\,m_{K^+}^4\,L^r_{7}
\nonumber\\&&
          - 256\,m_{\pi^0}^2\,m_{K^+}^2\,L^r_{8}
%%\nonumber\\&&
          - 768\,m_{\pi^0}^2\,m_{K^+}^2\,L^r_{7}
%\nonumber\\&&
          + 128\,m_{\pi^0}^4\,L^r_{8}
%\nonumber\\&&
          + 384\,m_{\pi^0}^4\,L^r_{7}
\nonumber\\&&
          - 4/3\,\overline{A}(m_{\pi^+}^2)\,m_{K^+}^2
%\nonumber\\&&
          + 16/3\,\overline{A}(m_{\pi^+}^2)\,m_{\pi^0}^2
%\nonumber\\&&
          - 2\,\overline{A}(m_{\pi^0}^2)\,m_{K^+}^2
%\nonumber\\&&
          + 2\,\overline{A}(m_{\pi^0}^2)\,m_{\pi^0}^2
\nonumber\\&&
          - \overline{A}(m_{K^+}^2)\,m_{\pi^0}^2
%\nonumber\\&&
          + 4/3\,\overline{A}(m_{K^0}^2)\,m_{K^+}^2
%\nonumber\\&&
          - 13/3\,\overline{A}(m_{K^0}^2)\,m_{\pi^0}^2
%\nonumber\\&&
          + 2\,\overline{A}(m_{\eta}^2)\,m_{K^+}^2
\nonumber\\&&
          - 2\,\overline{A}(m_{\eta}^2)\,m_{\pi^0}^2
%\nonumber\\&&
          \Big)
\nonumber\\&&
       + \frac{\sin\epsilon}{\sqrt{3}\,F_\pi^2} \, \Big(
%\nonumber\\&&
          + 6\,L^r_{9}t
%\nonumber\\&&
          - 1/4\,\overline{A}(m_{\pi^+}^2)
%\nonumber\\&&
          + 9/8\,\overline{A}(m_{\pi^0}^2)
%\nonumber\\&&
          + 7/4\,\overline{A}(m_{K^+}^2)
\nonumber\\&&
          + 3/2\,\overline{A}(m_{K^0}^2)
%\nonumber\\&&
          + 3/8\,\overline{A}(m_{\eta}^2)
\nonumber\\&&
          - 3\,\overline{B}_{22}(m_{\pi^+}^2,m_{K^0}^2,t)
%\nonumber\\&&
          - 9/2\,\overline{B}_{22}(m_{\pi^0}^2,m_{K^+}^2,t)
%\nonumber\\&&
          - 3/2\,\overline{B}_{22}(m_{K^+}^2,m_{\eta}^2,t)
%\nonumber\\&&
          \Big)
\nonumber\\&&
+\frac{1}{F_\pi^2}\Big(
       + 2\,L^r_{9}t
%\nonumber\\&&
          + 1/4\,\overline{A}(m_{\pi^+}^2)
%\nonumber\\&&
          + 1/8\,\overline{A}(m_{\pi^0}^2)
%\nonumber\\&&
          + 3/4\,\overline{A}(m_{K^+}^2)
%\nonumber\\&&
          + 3/8\,\overline{A}(m_{\eta}^2)
\nonumber\\&&
          - \overline{B}_{22}(m_{\pi^+}^2,m_{K^0}^2,t)
%\nonumber\\&&
          - 1/2\,\overline{B}_{22}(m_{\pi^0}^2,m_{K^+}^2,t)
%\nonumber\\&&
          - 3/2\,\overline{B}_{22}(m_{K^+}^2,m_{\eta}^2,t)
%\nonumber\\&&
         \Big)\,,
\nonumber\\
f_+^{K^0\pi^-(4)}(t) &=&
%\nonumber\\&&
       \frac{\sin\epsilon}{\sqrt{3} \,F_\pi^2}\,\Big(
%\nonumber\\&&
          + 3/4\,\overline{A}(m_{\pi^0}^2)
%\nonumber\\&&
          - 3/4\,\overline{A}(m_{\eta}^2)
%\nonumber\\&&
          - 3\,\overline{B}_{22}(m_{\pi^0}^2,m_{K^+}^2,t)
\nonumber\\&&
          + 3\,\overline{B}_{22}(m_{K^+}^2,m_{\eta}^2,t)
%\nonumber\\&&
          \Big)
\nonumber\\&&
+\frac{1}{F_\pi^2}\Big( 2\,L^r_{9}t
%\nonumber\\&&
          + 1/4\,\overline{A}(m_{\pi^+}^2)
%\nonumber\\&&
          + 1/8\,\overline{A}(m_{\pi^0}^2)
%\nonumber\\&&
          + 1/2\,\overline{A}(m_{K^+}^2)
%\nonumber\\&&
          + 1/4\,\overline{A}(m_{K^0}^2)
\nonumber\\&&
          + 3/8\,\overline{A}(m_{\eta}^2)
%\nonumber\\&&
          - \overline{B}_{22}(m_{\pi^+}^2,m_{K^0}^2,t)
%\nonumber\\&&
          - 1/2\,\overline{B}_{22}(m_{\pi^0}^2,m_{K^+}^2,t)
\nonumber\\&&
          - 3/2\,\overline{B}_{22}(m_{K^+}^2,m_{\eta}^2,t)
%\nonumber\\&&
         \Big)\,,
\nonumber\\
f_+^{K^+\pi^+(4)}(t) &=&
%\nonumber\\&&
       + \frac{\sin\epsilon}{\sqrt{3}F_\pi^2} \, \Big(
%\nonumber\\&&
          - 3/4\,\overline{A}(m_{\pi^0}^2)
%\nonumber\\&&
          + 3/4\,\overline{A}(m_{\eta}^2)
%\nonumber\\&&
          + 3\,\overline{B}_{22}(m_{\pi^0}^2,m_{K^0}^2,t)
\nonumber\\&&
          - 3\,\overline{B}_{22}(m_{K^0}^2,m_{\eta}^2,t)
%\nonumber\\&&
          \Big)
\nonumber\\&&
 +\frac{1}{F_\pi^2}\Big(        2\,L^r_{9}t
%\nonumber\\&&
          + 1/4\,\overline{A}(m_{\pi^+}^2)
%\nonumber\\&&
          + 1/8\,\overline{A}(m_{\pi^0}^2)
%\nonumber\\&&
          + 1/4\,\overline{A}(m_{K^+}^2)
%\nonumber\\&&
          + 1/2\,\overline{A}(m_{K^0}^2)
\nonumber\\&&
          + 3/8\,\overline{A}(m_{\eta}^2)
%\nonumber\\&&
          - \overline{B}_{22}(m_{\pi^+}^2,m_{K^+}^2,t)
%\nonumber\\&&
          - 1/2\,\overline{B}_{22}(m_{\pi^0}^2,m_{K^0}^2,t)
\nonumber\\&&
          - 3/2\,\overline{B}_{22}(m_{K^0}^2,m_{\eta}^2,t)
%\nonumber\\&&
         \Big)\,,
\nonumber\\
f_+^{K^0\pi^0(4)}(t) &=&
%\nonumber\\&&
       \frac{1}{F_\pi^2\, \left(m_{\pi^0}^2-m_{\eta}^2\right)} \, \Big(
%\nonumber\\&&
          + 2/3\,\overline{A}(m_{K^+}^2)\,m_{K^0}^2
%\nonumber\\&&
          + 1/3\,\overline{A}(m_{K^+}^2)\,m_{\pi^0}^2
\nonumber\\&&
          - 2/3\,\overline{A}(m_{K^0}^2)\,m_{K^0}^2
%\nonumber\\&&
          - 1/3\,\overline{A}(m_{K^0}^2)\,m_{\pi^0}^2
%\nonumber\\&&
          \Big)
\nonumber\\&&
       +\frac{ \sin\epsilon}{F_\pi^2\,\sqrt{3}\,
            \left(m_{\pi^0}^2-m_{\eta}^2\right)} \, \Big(
%\nonumber\\&&
          - 128\,m_{K^0}^4\,L^r_{8}%\nonumber\\&&
          - 384\,m_{K^0}^4\,L^r_{7}%\nonumber\\&&
          + 256\,m_{\pi^0}^2\,m_{K^0}^2\,L^r_{8}
\nonumber\\&&
          + 768\,m_{\pi^0}^2\,m_{K^0}^2\,L^r_{7}%\nonumber\\&&
          - 128\,m_{\pi^0}^4\,L^r_{8}%\nonumber\\&&
          - 384\,m_{\pi^0}^4\,L^r_{7}%\nonumber\\&&
          + 4/3\,\overline{A}(m_{\pi^+}^2)\,m_{K^0}^2
\nonumber\\&&
          - 16/3\,\overline{A}(m_{\pi^+}^2)\,m_{\pi^0}^2
%\nonumber\\&&
          + 2\,\overline{A}(m_{\pi^0}^2)\,m_{K^0}^2
%\nonumber\\&&
          - 2\,\overline{A}(m_{\pi^0}^2)\,m_{\pi^0}^2
%\nonumber\\&&
          - 4/3\,\overline{A}(m_{K^+}^2)\,m_{K^0}^2
\nonumber\\&&
          + 13/3\,\overline{A}(m_{K^+}^2)\,m_{\pi^0}^2
%\nonumber\\&&
          + \overline{A}(m_{K^0}^2)\,m_{\pi^0}^2
%\nonumber\\&&
          - 2\,\overline{A}(m_{\eta}^2)\,m_{K^0}^2
%\nonumber\\&&
          + 2\,\overline{A}(m_{\eta}^2)\,m_{\pi^0}^2
%\nonumber\\&&
          \Big)
\nonumber\\&&
       +\frac{ \sin\epsilon}{F_\pi^2\,\sqrt{3}} \, \Big(
%\nonumber\\&&
          - 6\,L^r_{9}t
%\nonumber\\&&
          + 1/4\,\overline{A}(m_{\pi^+}^2)
%\nonumber\\&&
          - 9/8\,\overline{A}(m_{\pi^0}^2)
%\nonumber\\&&
          - 3/2\,\overline{A}(m_{K^+}^2)
\nonumber\\&&
          - 7/4\,\overline{A}(m_{K^0}^2)
%\nonumber\\&&
          - 3/8\,\overline{A}(m_{\eta}^2)
%\nonumber\\&&
          + 3\,\overline{B}_{22}(m_{\pi^+}^2,m_{K^+}^2,t)
%\nonumber\\&&
          + 9/2\,\overline{B}_{22}(m_{\pi^0}^2,m_{K^0}^2,t)
\nonumber\\&&
          + 3/2\,\overline{B}_{22}(m_{K^0}^2,m_{\eta}^2,t)
%\nonumber\\&&
          \Big)
\nonumber\\&&
 +\frac{1}{F_\pi^2}\,\Big( 2\,L^r_{9}t
%\nonumber\\&&
          + 1/4\,\overline{A}(m_{\pi^+}^2)
%\nonumber\\&&
          + 1/8\,\overline{A}(m_{\pi^0}^2)
%\nonumber\\&&
          + 3/4\,\overline{A}(m_{K^0}^2)
%\nonumber\\&&
          + 3/8\,\overline{A}(m_{\eta}^2)
\nonumber\\&&
          - \overline{B}_{22}(m_{\pi^+}^2,m_{K^+}^2,t)
%\nonumber\\&&
          - 1/2\,\overline{B}_{22}(m_{\pi^0}^2,m_{K^0}^2,t)
%\nonumber\\&&
          - 3/2\,\overline{B}_{22}(m_{K^0}^2,m_{\eta}^2,t)
%\nonumber\\&&
         \Big)\,,;
\nonumber\\
f_-^{K^+\pi^0(4)}(t) &=& \frac{\sin\epsilon}{F_\pi^2\,\sqrt{3}} \, \Big(
%\nonumber\\&&
          - 6\,m_{K^+}^2\,L^r_{9}%\nonumber\\&&
          + 12\,m_{K^+}^2\,L^r_{5}%\nonumber\\&&
          + 6\,m_{\pi^0}^2\,L^r_{9}%\nonumber\\&&
          - 12\,m_{\pi^0}^2\,L^r_{5}%\nonumber\\&&
          + 1/2\,\overline{A}(m_{\pi^+}^2)
\nonumber\\&&
          + 3/4\,\overline{A}(m_{\pi^0}^2)
%\nonumber\\&&
          - 7/4\,\overline{A}(m_{K^+}^2)
%\nonumber\\&&
          - \overline{A}(m_{K^0}^2)
%\nonumber\\&&
          + 1/2\,\overline{A}(m_{\eta}^2)
\nonumber\\&&
          +\overline{B}(m_{\pi^+}^2,m_{K^0}^2,t)\,(
          - 1/2\,\,t
%\nonumber\\&&
          - 3/2\,m_{K^+}^2
%\nonumber\\&&
          + 5/2\,m_{\pi^0}^2)
\nonumber\\&&
          + \overline{B}(m_{\pi^0}^2,m_{K^+}^2,t)\,(
          - 3/4\,t
%\nonumber\\&&
          - 5/4\,m_{K^+}^2
%\nonumber\\&&
          + 15/4\,m_{\pi^0}^2
\nonumber\\&&
          +\,\overline{B}(m_{K^+}^2,m_{\eta}^2,t)\,(
          + 1/4\,t
%\nonumber\\&&
          - 5/4\,\,m_{K^+}^2
%\nonumber\\&&
          + 3/4\,\,m_{\pi^0}^2)
\nonumber\\&&
          + \overline{B}_{1}(m_{\pi^+}^2,m_{K^0}^2,t)\,(
          + 1/2\,t
%\nonumber\\&&
          + 9/2\,m_{K^+}^2
%\nonumber\\&&
          - 13/2\,m_{\pi^0}^2)
\nonumber\\&&
          +\,\overline{B}_{1}(m_{\pi^0}^2,m_{K^+}^2,t)\,(
          + 3/4\,t
%\nonumber\\&&
          + 19/4\,m_{K^+}^2
%\nonumber\\&&
          - 39/4\,m_{\pi^0}^2)
\nonumber\\&&
          +\,\overline{B}_{1}(m_{K^+}^2,m_{\eta}^2,t)(
          - 1/4\,t
%\nonumber\\&&
          + 13/4\,m_{K^+}^2
%\nonumber\\&&
          - 9/4\,m_{\pi^0}^2)
\nonumber\\&&
          + \overline{B}_{21}(m_{\pi^+}^2,m_{K^0}^2,t)\,(
          +\,t
%\nonumber\\&&
          - 3\,m_{K^+}^2
%\nonumber\\&&
          + 3\,m_{\pi^0}^2)
\nonumber\\&&
          +\,\overline{B}_{21}(m_{\pi^0}^2,m_{K^+}^2,t)\,(
          + 3/2\,t
%\nonumber\\&&
          - 9/2\,m_{K^+}^2
%\nonumber\\&&
          + 9/2\,m_{\pi^0}^2)
\nonumber\\&&
          +\,\overline{B}_{21}(m_{K^+}^2,m_{\eta}^2,t)\,(
          - 1/2\,t
%\nonumber\\&&
          - 3/2\,m_{K^+}^2
%\nonumber\\&&
          + 3/2\,m_{\pi^0}^2)
\nonumber\\&&
          + \overline{B}_{22}(m_{\pi^+}^2,m_{K^0}^2,t)
%\nonumber\\&&
          + 3/2\,\overline{B}_{22}(m_{\pi^0}^2,m_{K^+}^2,t)
%\nonumber\\&&
          - 1/2\,\overline{B}_{22}(m_{K^+}^2,m_{\eta}^2,t)
%\nonumber\\&&
          \Big)
\nonumber\\&&
+\frac{1}{F_\pi^2}\,\Big(
       - 2\,m_{K^+}^2\,L^r_{9}%\nonumber\\&&
          + 4\,m_{K^+}^2\,L^r_{5}%\nonumber\\&&
          + 2\,m_{\pi^0}^2\,L^r_{9}%\nonumber\\&&
          - 4\,m_{\pi^0}^2\,L^r_{5}%\nonumber\\&&
          - 1/2\,\overline{A}(m_{\pi^+}^2)
\nonumber\\&&
          + 1/12\,\overline{A}(m_{\pi^0}^2)
%\nonumber\\&&
          - 5/12\,\overline{A}(m_{K^+}^2)
%\nonumber\\&&
          + \overline{A}(m_{K^0}^2)
%\nonumber\\&&
          + 1/2\,\overline{A}(m_{\eta}^2)
\nonumber\\&&
          +\,\overline{B}(m_{\pi^+}^2,m_{K^0}^2,t)\,(
          + 1/2\,t
%\nonumber\\&&
          - 1/2\,m_{K^+}^2
%\nonumber\\&&
          - 1/2\,m_{\pi^0}^2)
\nonumber\\&&
          +\,\overline{B}(m_{\pi^0}^2,m_{K^+}^2,t)\,(
          - 1/12\,t
%\nonumber\\&&
          + 1/12\,m_{K^+}^2
%\nonumber\\&&
          + 5/12\,m_{\pi^0}^2)
\nonumber\\&&
          +\,\overline{B}(m_{K^+}^2,m_{\eta}^2,t)\,(
          + 1/4\,t
%\nonumber\\&&
          - 7/12\,m_{K^+}^2
%\nonumber\\&&
          + 1/12\,m_{\pi^0}^2)
\nonumber\\&&
+\,\overline{B}_{1}(m_{\pi^+}^2,m_{K^0}^2,t)\,(
          - 1/2\,t
%\nonumber\\&&
          + 3/2\,m_{K^+}^2
%\nonumber\\&&
          + 1/2\,m_{\pi^0}^2)
\nonumber\\&&
+\,\overline{B}_{1}(m_{\pi^0}^2,m_{K^+}^2,t)\,(
          + 1/12\,t
%\nonumber\\&&
          + 1/12\,m_{K^+}^2
%\nonumber\\&&
          - 13/12\,m_{\pi^0}^2)
\nonumber\\&&
+\,\overline{B}_{1}(m_{K^+}^2,m_{\eta}^2,t)\,(
          - 1/4\,t
%\nonumber\\&&
          + 23/12\,m_{K^+}^2
%\nonumber\\&&
          - 11/12\,m_{\pi^0}^2)
\nonumber\\&&
+ \overline{B}_{21}(m_{\pi^+}^2,m_{K^0}^2,t)\,(
          -\,t
%\nonumber\\&&
          -\,m_{K^+}^2
%\nonumber\\&&
          +\,m_{\pi^0}^2)
\nonumber\\&&
+\,\overline{B}_{21}(m_{\pi^0}^2,m_{K^+}^2,t)\,(
          + 1/6\,t
%\nonumber\\&&
          - 1/2\,m_{K^+}^2
%\nonumber\\&&
          + 1/2\,m_{\pi^0}^2)
\nonumber\\&&
\,\overline{B}_{21}(m_{K^+}^2,m_{\eta}^2,t)\,(
          - 1/2\,t
%\nonumber\\&&
          - 3/2\,m_{K^+}^2
%\nonumber\\&&
          + 3/2\,m_{\pi^0}^2)
\nonumber\\&&
          - \overline{B}_{22}(m_{\pi^+}^2,m_{K^0}^2,t)
%\nonumber\\&&
          + 1/6\,\overline{B}_{22}(m_{\pi^0}^2,m_{K^+}^2,t)
%\nonumber\\&&
          - 1/2\,\overline{B}_{22}(m_{K^+}^2,m_{\eta}^2,t)
%\nonumber\\&&
         \Big)\,,
\nonumber\\
f_-^{K^0\pi^-(4)}(t) &=& \frac{\sin\epsilon}{F_\pi^2\,\sqrt{3}} \, \Big(
%\nonumber\\&&
          - 1/2\,\overline{A}(m_{\pi^0}^2)
%\nonumber\\&&
          + 1/2\,\overline{A}(m_{K^+}^2)
%\nonumber\\&&
          + \overline{A}(m_{\eta}^2)
\nonumber\\&&
          + \overline{B}(m_{\pi^0}^2,m_{K^+}^2,t)\,(
          + 1/2\,t
%\nonumber\\&&
          - 3/2\,m_{K^0}^2
%\nonumber\\&&
          + 1/2\,m_{\pi^+}^2)
\nonumber\\&&
          + \overline{B}(m_{K^+}^2,m_{\eta}^2,t)\,(
          + 1/2\,t
%\nonumber\\&&
          + 1/2\,m_{K^0}^2
%\nonumber\\&&
          - 1/2\,m_{\pi^0}^2
%\nonumber\\&&
          -\,m_{\pi^+}^2)
\nonumber\\&&
          +\overline{B}_{1}(m_{\pi^0}^2,m_{K^+}^2,t)\,(
          - 1/2\,t
%\nonumber\\&&
          + 9/2\,m_{K^0}^2
%\nonumber\\&&
          - 5/2\,m_{\pi^+}^2)
\nonumber\\&&
          +\overline{B}_{1}(m_{K^+}^2,m_{\eta}^2,t)\,(
          - 1/2\,t
%\nonumber\\&&
          - 5/2\,m_{K^0}^2
%\nonumber\\&&
          + \,m_{\pi^0}^2
%\nonumber\\&&
          + 7/2\,m_{\pi^+}^2)
\nonumber\\&&
          + \overline{B}_{21}(m_{\pi^0}^2,m_{K^+}^2,t)\,(
          - \,t
%\nonumber\\&&
          - 3\,m_{K^0}^2
%\nonumber\\&&
          + 3\,m_{\pi^+}^2)
\nonumber\\&&
          + \overline{B}_{21}(m_{K^+}^2,m_{\eta}^2,t)\,(
          - \,t
%\nonumber\\&&
          + 3\,m_{K^0}^2
%\nonumber\\&&
          - 3\,m_{\pi^+}^2)
\nonumber\\&&
          - \overline{B}_{22}(m_{\pi^0}^2,m_{K^+}^2,t)
%\nonumber\\&&
          - \overline{B}_{22}(m_{K^+}^2,m_{\eta}^2,t)
%\nonumber\\&&
          \Big)
\nonumber\\&&
 +\frac{1}{F_\pi^2}\,\Big(
       - 2\,m_{K^0}^2\,L^r_{9}%\nonumber\\&&
          + 4\,m_{K^0}^2\,L^r_{5}%\nonumber\\&&
          - 4\,m_{\pi^0}^2\,L^r_{5}%\nonumber\\&&
          + 2\,m_{\pi^+}^2\,L^r_{9}%\nonumber\\&&
          - 1/6\,\overline{A}(m_{\pi^+}^2)
\nonumber\\&&
          - 1/4\,\overline{A}(m_{\pi^0}^2)
%\nonumber\\&&
          + 1/4\,\overline{A}(m_{K^+}^2)
%\nonumber\\&&
          + 1/3\,\overline{A}(m_{K^0}^2)
%\nonumber\\&&
          + 1/2\,\overline{A}(m_{\eta}^2)
\nonumber\\&&
          + \overline{B}(m_{\pi^+}^2,m_{K^0}^2,t)\,(
          + 1/6\,t
%\nonumber\\&&
          - 1/6\,m_{K^0}^2
%\nonumber\\&&
          + 1/6\,m_{\pi^+}^2)
\nonumber\\&&
          + \overline{B}(m_{\pi^0}^2,m_{K^+}^2,t)\,(
          + 1/4\,t
%\nonumber\\&&
          - 1/4\,m_{K^0}^2
%\nonumber\\&&
          - 1/4\,m_{\pi^0}^2)
\nonumber\\&&
          + \overline{B}(m_{K^+}^2,m_{\eta}^2,t)\,(
          + 1/4\,t
%\nonumber\\&&
          - 7/12\,m_{K^0}^2
%\nonumber\\&&
          - 1/6\,m_{\pi^0}^2
%\nonumber\\&&
          + 1/4\,m_{\pi^+}^2)
\nonumber\\&&
          +\,\overline{B}_{1}(m_{\pi^+}^2,m_{K^0}^2,t)\,(
          - 1/6\,t
%\nonumber\\&&
          + 5/6\,m_{K^0}^2
%\nonumber\\&&
          - 5/6\,m_{\pi^+}^2)
\nonumber\\&&
          +\,\overline{B}_{1}(m_{\pi^0}^2,m_{K^+}^2,t)\,(
          - 1/4\,t
%\nonumber\\&&
          + 3/4\,m_{K^0}^2
%\nonumber\\&&
          + 1/2\,m_{\pi^0}^2
%\nonumber\\&&
          - 1/4\,m_{\pi^+}^2)
\nonumber\\&&
          +\,\overline{B}_{1}(m_{K^+}^2,m_{\eta}^2,t)\,(
          - 1/4\,t
%\nonumber\\&&
          + 23/12\,m_{K^0}^2
%\nonumber\\&&
          + 1/3\,m_{\pi^0}^2
%\nonumber\\&&
          - 5/4\,m_{\pi^+}^2)
\nonumber\\&&
          +\,\overline{B}_{21}(m_{\pi^+}^2,m_{K^0}^2,t)\,(
          - 1/3\,t
%\nonumber\\&&
          - \,m_{K^0}^2
%\nonumber\\&&
          + \,m_{\pi^+}^2)
\nonumber\\&&
          +\,\overline{B}_{21}(m_{\pi^0}^2,m_{K^+}^2,t)\,(
          - 1/2\,t
%\nonumber\\&&
          - 1/2\,m_{K^0}^2
%\nonumber\\&&
          + 1/2\,m_{\pi^+}^2)
\nonumber\\&&
          +\overline{B}_{21}(m_{K^+}^2,m_{\eta}^2,t)\,(
          - 1/2\,t
%\nonumber\\&&
          - 3/2\,m_{K^0}^2
%\nonumber\\&&
          + 3/2\,m_{\pi^+}^2)
\nonumber\\&&
          - 1/3\,\overline{B}_{22}(m_{\pi^+}^2,m_{K^0}^2,t)
%\nonumber\\&&
          - 1/2\,\overline{B}_{22}(m_{\pi^0}^2,m_{K^+}^2,t)
%\nonumber\\&&
          - 1/2\,\overline{B}_{22}(m_{K^+}^2,m_{\eta}^2,t)
%\nonumber\\&&
         \Big)\,,
\nonumber\\
f_-^{K^+\pi^+(4)}(t) &=& \frac{\sin\epsilon}{F_\pi^2\,\sqrt{3}} \, \Big(
%\nonumber\\&&
          + 1/2\,\overline{A}(m_{\pi^0}^2)
%\nonumber\\&&
          - 1/2\,\overline{A}(m_{K^0}^2)
%\nonumber\\&&
          - \overline{A}(m_{\eta}^2)
\nonumber\\&&
          +\,\overline{B}(m_{\pi^0}^2,m_{K^0}^2,t)\,(
          - 1/2\,t
%\nonumber\\&&
          + 3/2\,m_{K^+}^2
%\nonumber\\&&
          - 1/2\,m_{\pi^+}^2)
\nonumber\\&&
          +\,\overline{B}(m_{K^0}^2,m_{\eta}^2,t)\,(
          - 1/2\,t
%\nonumber\\&&
          - 1/2\,m_{K^+}^2
%\nonumber\\&&
          + 1/2\,m_{\pi^0}^2
%\nonumber\\&&
          + \,m_{\pi^+}^2)
\nonumber\\&&
          + \,\overline{B}_{1}(m_{\pi^0}^2,m_{K^0}^2,t)\,(
          + 1/2\,t
%\nonumber\\&&
          - 9/2\,m_{K^+}^2
%\nonumber\\&&
          + 5/2\,m_{\pi^+}^2)
\nonumber\\&&
          + ,\overline{B}_{1}(m_{K^0}^2,m_{\eta}^2,t)\,(
          + 1/2\,t
%\nonumber\\&&
          + 5/2\,m_{K^+}^2
%\nonumber\\&&
          - \,m_{\pi^0}^2
%\nonumber\\&&
          - \,m_{\pi^+}^2)
\nonumber\\&&
          + \overline{B}_{21}(m_{\pi^0}^2,m_{K^0}^2,t)\,(
          + \,t
%\nonumber\\&&
          + 3\,m_{K^+}^2
%\nonumber\\&&
          - 3\,m_{\pi^+}^2)
\nonumber\\&&
          + \overline{B}_{21}(m_{K^0}^2,m_{\eta}^2,t)\,(
          + \,t
%\nonumber\\&&
          - 3\,m_{K^+}^2
%\nonumber\\&&
          + 3\,m_{\pi^+}^2)
\nonumber\\&&
          + \overline{B}_{22}(m_{\pi^0}^2,m_{K^0}^2,t)
%\nonumber\\&&
          + \overline{B}_{22}(m_{K^0}^2,m_{\eta}^2,t)
%\nonumber\\&&
          \Big)
\nonumber\\&&
 +\frac{1}{F_\pi^2}\,\Big(
       - 2\,m_{K^+}^2\,L^r_{9}%\nonumber\\&&
          + 4\,m_{K^+}^2\,L^r_{5}%\nonumber\\&&
          - 4\,m_{\pi^0}^2\,L^r_{5}%\nonumber\\&&
          + 2\,m_{\pi^+}^2\,L^r_{9}%\nonumber\\&&
          - 1/6\,\overline{A}(m_{\pi^+}^2)
\nonumber\\&&
          - 1/4\,\overline{A}(m_{\pi^0}^2)
%\nonumber\\&&
          + 1/3\,\overline{A}(m_{K^+}^2)
%\nonumber\\&&
          + 1/4\,\overline{A}(m_{K^0}^2)
%\nonumber\\&&
          + 1/2\,\overline{A}(m_{\eta}^2)
\nonumber\\&&
          + 1/6\,\overline{B}(m_{\pi^+}^2,m_{K^+}^2,t)\,(
          + 1/6\,t
%\nonumber\\&&
          - 1/6\,m_{K^+}^2
%\nonumber\\&&
          + 1/6\,m_{\pi^+}^2)
\nonumber\\&&
          + 1/4\,\overline{B}(m_{\pi^0}^2,m_{K^0}^2,t)\,(
          + 1/4\,t
%\nonumber\\&&
          - 1/4\,m_{K^+}^2
%\nonumber\\&&
          - 1/4\,m_{\pi^0}^2)
\nonumber\\&&
          + \,\overline{B}(m_{K^0}^2,m_{\eta}^2,t)\,(
          + 1/4\,t
%\nonumber\\&&
          - 7/12\,m_{K^+}^2
%\nonumber\\&&
          - 1/6\,m_{\pi^0}^2
%\nonumber\\&&
          + 1/4\,m_{\pi^+}^2)
\nonumber\\&&
          +\,\overline{B}_{1}(m_{\pi^+}^2,m_{K^+}^2,t)\,(
          - 1/6\,t
%\nonumber\\&&
          + 5/6\,m_{K^+}^2
%\nonumber\\&&
          - 5/6\,m_{\pi^+}^2)
\nonumber\\&&
          +\,\overline{B}_{1}(m_{\pi^0}^2,m_{K^0}^2,t)\,(
          - 1/4\,t
%\nonumber\\&&
          + 3/4\,m_{K^+}^2
%\nonumber\\&&
          + 1/2\,m_{\pi^0}^2
%\nonumber\\&&
          - 1/4\,m_{\pi^+}^2)
\nonumber\\&&
          +\,\overline{B}_{1}(m_{K^0}^2,m_{\eta}^2,t)\,(
          - 1/4\,t
%\nonumber\\&&
          + 23/12\,m_{K^+}^2
%\nonumber\\&&
          + 1/3\,m_{\pi^0}^2
%\nonumber\\&&
          - 5/4\,m_{\pi^+}^2)
\nonumber\\&&
          +\,\overline{B}_{21}(m_{\pi^+}^2,m_{K^+}^2,t)\,(
          - 1/3\,t
%\nonumber\\&&
          - \,m_{K^+}^2
%\nonumber\\&&
          + \,m_{\pi^+}^2)
\nonumber\\&&
          +\,\overline{B}_{21}(m_{\pi^0}^2,m_{K^0}^2,t)\,(
          - 1/2\,t
%\nonumber\\&&
          - 1/2\,m_{K^+}^2
%\nonumber\\&&
          + 1/2\,m_{\pi^+}^2)
\nonumber\\&&
          +\,\overline{B}_{21}(m_{K^0}^2,m_{\eta}^2,t)\,(
          - 1/2\,t
%\nonumber\\&&
          - 3/2\,m_{K^+}^2
%\nonumber\\&&
          + 3/2\,m_{\pi^+}^2)
\nonumber\\&&
          - 1/3\,\overline{B}_{22}(m_{\pi^+}^2,m_{K^+}^2,t)
%\nonumber\\&&
          - 1/2\,\overline{B}_{22}(m_{\pi^0}^2,m_{K^0}^2,t)
%\nonumber\\&&
          - 1/2\,\overline{B}_{22}(m_{K^0}^2,m_{\eta}^2,t)
%\nonumber\\&&
         \Big)
\nonumber\\
f_-^{K^0\pi^0(4)}(t) &=& \frac{\sin\epsilon}{F_\pi^2\,\sqrt{3}} \, \Big(
%\nonumber\\&&
          + 6\,m_{K^0}^2\,L^r_{9}%\nonumber\\&&
          - 12\,m_{K^0}^2\,L^r_{5}%\nonumber\\&&
          - 6\,m_{\pi^0}^2\,L^r_{9}%\nonumber\\&&
          + 12\,m_{\pi^0}^2\,L^r_{5}%\nonumber\\&&
          - 1/2\,\overline{A}(m_{\pi^+}^2)
\nonumber\\&&
          - 3/4\,\overline{A}(m_{\pi^0}^2)
%\nonumber\\&&
          + \overline{A}(m_{K^+}^2)
%\nonumber\\&&
          + 7/4\,\overline{A}(m_{K^0}^2)
%\nonumber\\&&
          - 1/2\,\overline{A}(m_{\eta}^2)
\nonumber\\&&
          + \,\overline{B}(m_{\pi^+}^2,m_{K^+}^2,t)\,(
          + 1/2\,t
%\nonumber\\&&
          + 3/2\,m_{K^0}^2
%\nonumber\\&&
          - 5/2\,m_{\pi^0}^2)
\nonumber\\&&
          + \,\overline{B}(m_{\pi^0}^2,m_{K^0}^2,t)\,(
          + 3/4\,t
%\nonumber\\&&
          + 5/4\,m_{K^0}^2
%\nonumber\\&&
          - 15/4\,m_{\pi^0}^2)
\nonumber\\&&
          +\,\overline{B}(m_{K^0}^2,m_{\eta}^2,t)\,(
          - 1/4\,t
%\nonumber\\&&
          + 5/4\,m_{K^0}^2
%\nonumber\\&&
          - 3/4\,m_{\pi^0}^2)
\nonumber\\&&
          +\,\overline{B}_{1}(m_{\pi^+}^2,m_{K^+}^2,t)\,(
          - 1/2\,t
%\nonumber\\&&
          - 9/2\,m_{K^0}^2
%\nonumber\\&&
          + 13/2\,m_{\pi^0}^2)
\nonumber\\&&
          +\,\overline{B}_{1}(m_{\pi^0}^2,m_{K^0}^2,t)\,(
          - 3/4\,t
%\nonumber\\&&
          - 19/4\,m_{K^0}^2
%\nonumber\\&&
          + 39/4\,m_{\pi^0}^2)
\nonumber\\&&
          +\,\overline{B}_{1}(m_{K^0}^2,m_{\eta}^2,t)\,(
          + 1/4\,t
%\nonumber\\&&
          - 13/4\,m_{K^0}^2
%\nonumber\\&&
          + 9/4\,m_{\pi^0}^2)
\nonumber\\&&
          + \overline{B}_{21}(m_{\pi^+}^2,m_{K^+}^2,t)\,(
          - \,t
%\nonumber\\&&
          + 3\,m_{K^0}^2
%\nonumber\\&&
          - 3\,m_{\pi^0}^2)
\nonumber\\&&
          +\,\overline{B}_{21}(m_{\pi^0}^2,m_{K^0}^2,t)\,(
          - 3/2\,t
%\nonumber\\&&
          + 9/2\,m_{K^0}^2
%\nonumber\\&&
          - 9/2\,m_{\pi^0}^2)
\nonumber\\&&
          +\,\overline{B}_{21}(m_{K^0}^2,m_{\eta}^2,t)\,(
          + 1/2\,t
%\nonumber\\&&
          + 3/2\,m_{K^0}^2
%\nonumber\\&&
          - 3/2\,m_{\pi^0}^2)
\nonumber\\&&
          - \overline{B}_{22}(m_{\pi^+}^2,m_{K^+}^2,t)
%\nonumber\\&&
          - 3/2\,\overline{B}_{22}(m_{\pi^0}^2,m_{K^0}^2,t)
%\nonumber\\&&
          + 1/2\,\overline{B}_{22}(m_{K^0}^2,m_{\eta}^2,t)
%\nonumber\\&&
          \Big)
\nonumber\\&&
+\frac{1}{F_\pi^2}\,\Big(
       - 2\,m_{K^0}^2\,L^r_{9}%\nonumber\\&&
          + 4\,m_{K^0}^2\,L^r_{5}%\nonumber\\&&
          + 2\,m_{\pi^0}^2\,L^r_{9}%\nonumber\\&&
          - 4\,m_{\pi^0}^2\,L^r_{5}%\nonumber\\&&
          - 1/2\,\overline{A}(m_{\pi^+}^2)
\nonumber\\&&
          + 1/12\,\overline{A}(m_{\pi^0}^2)
%\nonumber\\&&
          + \overline{A}(m_{K^+}^2)
%\nonumber\\&&
          - 5/12\,\overline{A}(m_{K^0}^2)
%\nonumber\\&&
          + 1/2\,\overline{A}(m_{\eta}^2)
\nonumber\\&&
          + ,\overline{B}(m_{\pi^+}^2,m_{K^+}^2,t)\,(
          + 1/2\,t
%\nonumber\\&&
          - 1/2\,m_{K^0}^2
%\nonumber\\&&
          - 1/2\,m_{\pi^0}^2)
\nonumber\\&&
          +\,\overline{B}(m_{\pi^0}^2,m_{K^0}^2,t)\,(
          - 1/12\,t
%\nonumber\\&&
          + 1/12\,m_{K^0}^2
%\nonumber\\&&
          + 5/12\,m_{\pi^0}^2)
\nonumber\\&&
          +,\overline{B}(m_{K^0}^2,m_{\eta}^2,t)\,(
          + 1/4\,t
%\nonumber\\&&
          - 7/12\,m_{K^0}^2
%\nonumber\\&&
          + 1/12\,m_{\pi^0}^2)
\nonumber\\&&
          +\,\overline{B}_{1}(m_{\pi^+}^2,m_{K^+}^2,t)\,(
          - 1/2\,t
%\nonumber\\&&
          + 3/2\,m_{K^0}^2
%\nonumber\\&&
          + 1/2\,m_{\pi^0}^2)
\nonumber\\&&
          +\,\overline{B}_{1}(m_{\pi^0}^2,m_{K^0}^2,t)\,(
          + 1/12\,t
%\nonumber\\&&
          + 1/12\,m_{K^0}^2
%\nonumber\\&&
          - 13/12\,m_{\pi^0}^2)
\nonumber\\&&
          +\,\overline{B}_{1}(m_{K^0}^2,m_{\eta}^2,t)\,(
          - 1/4\,t
%\nonumber\\&&
          + 23/12\,m_{K^0}^2
%\nonumber\\&&
          - 11/12\,m_{\pi^0}^2)
\nonumber\\&&
          + \overline{B}_{21}(m_{\pi^+}^2,m_{K^+}^2,t)\,(
          - \,t
%\nonumber\\&&
          - \,m_{K^0}^2
%\nonumber\\&&
          + \,m_{\pi^0}^2)
\nonumber\\&&
          + \,\overline{B}_{21}(m_{\pi^0}^2,m_{K^0}^2,t)\,(
          + 1/6\,t
%\nonumber\\&&
          - 1/2\,m_{K^0}^2
%\nonumber\\&&
          + 1/2\,m_{\pi^0}^2)
\nonumber\\&&
          +\,\overline{B}_{21}(m_{K^0}^2,m_{\eta}^2,t)\,(
          - 1/2\,t
%\nonumber\\&&
          - 3/2\,m_{K^0}^2
%\nonumber\\&&
          + 3/2\,m_{\pi^0}^2)
\nonumber\\&&
          - \overline{B}_{22}(m_{\pi^+}^2,m_{K^+}^2,t)
%\nonumber\\&&
          + 1/6\,\overline{B}_{22}(m_{\pi^0}^2,m_{K^0}^2,t)
%\nonumber\\&&
          - 1/2\,\overline{B}_{22}(m_{K^0}^2,m_{\eta}^2,t)
\Big)\,.
\ea

\section{The order $p^6$ LECs dependent part}
\label{App:Ci}

In this appendix we write out explicitly the part dependent on the
order $p^6$ LECs $C_i^r$.
We use here a notation which uses the property (\ref{expansion}).
\ba
\left.f_\ell^{K^+\pi^0}(t)\right|_{C_i^r} &=&
\frac{1}{F_\pi^4}\left(f_\ell^A(t)
+\frac{\sin\epsilon}{\sqrt{3}}f_\ell^B(t)
+\frac{\sin\epsilon}{\sqrt{3}\left(m_{\pi^0}^2-m_\eta^2\right)}f_\ell^E(t)
\right)\,,
\nonumber\\
\left.f_\ell^{K^0\pi^-}(t)\right|_{C_i^r} &=&
\frac{1}{F_\pi^4}\left(f_\ell^A(t)
-\frac{\sin\epsilon}{\sqrt{3}}f_\ell^D(t)
\right)\,,
\nonumber\\
\left.f_\ell^{K^+\pi^+}(t)\right|_{C_i^r} &=&
\frac{1}{F_\pi^4}\left(f_\ell^A(t)
+\frac{\sin\epsilon}{\sqrt{3}}f_\ell^D(t)
\right)\,,
\nonumber\\
\left.f_\ell^{K^0\pi^0}(t)\right|_{C_i^r} &=&
\frac{1}{F_\pi^4}\left(f_\ell^A(t)
-\frac{\sin\epsilon}{\sqrt{3}}f_\ell^B(t)
-\frac{\sin\epsilon}{\sqrt{3}\left(m_{\pi^0}^2-m_\eta^2\right)}f_\ell^E(t)
\right)\,.
\ea
We also use the notation
\be
m_\sigma^2 = m_{K^+}^2+m_{K^0}^2-m_\pi^2\,.
\ee
The pion mass we have used generically since they are the same to the order
of our calculation.

The $C_i^r$ dependence is now given by
\ba
f_+^A(t)&=&
       + t^2\,(  - 4\,C^r_{88} + 4\,C^r_{90} )
%\nonumber\\&&
       + m_\sigma^2\,t\,(  - 4\,C^r_{12} - 16\,C^r_{13} - 4\,C^r_{63} - 4\,C^r_{64} - 2\,C^r_{90} )
\nonumber\\&&
       + m_\pi^2\,t\,(  - 12\,C^r_{12} - 32\,C^r_{13} - 4\,C^r_{63} - 8\,C^r_{64} - 4\,C^r_{65} - 6\,C^r_{90}
          )
%\nonumber\\&&
       + m_\sigma^4\,(  - 2\,C^r_{12} - 2\,C^r_{34} )
\nonumber\\&&
       + m_\pi^2\,m_\sigma^2\,( 4\,C^r_{12} + 4\,C^r_{34} )
%\nonumber\\&&
       + m_\pi^4\,(  - 2\,C^r_{12} - 2\,C^r_{34} )\,,
\nonumber\\
f_+^B(t) &=&
       + t^2\,(  - 12\,C^r_{88} + 12\,C^r_{90} )
%\nonumber\\&&
       + m_\sigma^2\,t\,(  - 4\,C^r_{12} - 48\,C^r_{13} - 4\,C^r_{63} - 12\,C^r_{64} - 2\,C^r_{90} )
\nonumber\\&&
       + m_\pi^2\,t\,(  - 44\,C^r_{12} - 96\,C^r_{13} - 20\,C^r_{63} - 24\,C^r_{64} - 12\,C^r_{65} - 22\,
         C^r_{90} )
\nonumber\\&&
       + m_\sigma^4\,( 2\,C^r_{12} + 16\,C^r_{14} + 16\,C^r_{17} + 48\,C^r_{18} - 14\,C^r_{34} - 24\,
         C^r_{35} )
\nonumber\\&&
       + m_\pi^2\,m_\sigma^2\,(  - 4\,C^r_{12} - 32\,C^r_{14} - 32\,C^r_{17} - 96\,C^r_{18} + 28\,C^r_{34}
          + 48\,C^r_{35} )
\nonumber\\&&
       + m_\pi^4\,( 2\,C^r_{12} + 16\,C^r_{14} + 16\,C^r_{17} + 48\,C^r_{18} - 14\,C^r_{34} - 24\,C^r_{35} )\,,
\nonumber\\
f_+^E(t) &=&
       + m_\sigma^6\,( 96\,C^r_{19} + 64\,C^r_{20} + 64\,C^r_{31} + 64\,C^r_{32} + 128\,C^r_{33} )
\nonumber\\&&
       + m_\pi^2\,m_\sigma^4\,(  - 32\,C^r_{14} - 32\,C^r_{17} - 96\,C^r_{18} )
%\nonumber\\&&
       + m_\pi^4\,m_\sigma^2\,( 64\,C^r_{14} + 64\,C^r_{17} + 192\,C^r_{18}
\nonumber\\&&
 - 288\,C^r_{19} - 192\,
         C^r_{20} - 192\,C^r_{31} - 192\,C^r_{32} - 384\,C^r_{33} )
\nonumber\\&&
       + m_\pi^6\,(  - 32\,C^r_{14} - 32\,C^r_{17} - 96\,C^r_{18} + 192\,C^r_{19} + 128\,C^r_{20} + 128
%\nonumber\\&&
         \,C^r_{31} + 128\,C^r_{32}
\nonumber\\&& + 256\,C^r_{33} )\,,
\nonumber\\
f_+^D(t) &=&
       + m_\sigma^2\,t\,( 8\,C^r_{12} - 8\,C^r_{63} + 8\,C^r_{65} + 4\,C^r_{90} )
%\nonumber\\&&
       + m_\pi^2\,t\,(  - 8\,C^r_{12} + 8\,C^r_{63} - 8\,C^r_{65} - 4\,C^r_{90} )
\nonumber\\&&
       + m_\sigma^4\,( 8\,C^r_{12} + 8\,C^r_{34} )
%\nonumber\\&&
       + m_\pi^2\,m_\sigma^2\,(  - 16\,C^r_{12} - 16\,C^r_{34} )
%\nonumber\\&&
       + m_\pi^4\,( 8\,C^r_{12} + 8\,C^r_{34} )\,,
\nonumber\\
f_-^A(t) &=&
       + m_\sigma^2\,t\,(  - 4\,C^r_{12} + 2\,C^r_{88} - 2\,C^r_{90} )
%\nonumber\\&&
       + m_\pi^2\,t\,( 4\,C^r_{12} - 2\,C^r_{88} + 2\,C^r_{90} )
\nonumber\\&&
       + m_\sigma^4\,( 6\,C^r_{12} + 8\,C^r_{13} + 4\,C^r_{14} + 4\,C^r_{15} + 2\,C^r_{34} + 2\,C^r_{63} +
%\nonumber\\&&
         2\,C^r_{64} + C^r_{90} )
\nonumber\\&&
       + m_\pi^2\,m_\sigma^2\,( 12\,C^r_{12} + 8\,C^r_{13} + 4\,C^r_{15} + 8\,C^r_{17} + 4\,C^r_{34} + 2\,
%\nonumber\\&&
         C^r_{64} + 2\,C^r_{65} + 2\,C^r_{90} )
\nonumber\\&&
       + m_\pi^4\,(  - 18\,C^r_{12} - 16\,C^r_{13} - 4\,C^r_{14} - 8\,C^r_{15} - 8\,C^r_{17} - 6\,C^r_{34}
%\nonumber\\&&
          - 2\,C^r_{63} - 4\,C^r_{64}
\nonumber\\&&
 - 2\,C^r_{65} - 3\,C^r_{90} )\,,
\nonumber\\
f_-^B(t) &=&
       + m_\sigma^2\,t\,(  - 4\,C^r_{12} + 2\,C^r_{88} - 2\,C^r_{90} )
%\nonumber\\&&
       + m_\pi^2\,t\,( 4\,C^r_{12} - 2\,C^r_{88} + 2\,C^r_{90} )
\nonumber\\&&
       + m_\sigma^4\,(  - 6\,C^r_{12} + 8\,C^r_{13} - 4\,C^r_{14} + 4\,C^r_{15} - 32\,C^r_{17} - 48\,
%\nonumber\\&&
         C^r_{18} - 18\,C^r_{34} - 24\,C^r_{35} - 2\,C^r_{63}
\nonumber\\&&
 + 2\,C^r_{64} - C^r_{90} )
\nonumber\\&&
       + m_\pi^2\,m_\sigma^2\,( 36\,C^r_{12} + 8\,C^r_{13} + 16\,C^r_{14} + 4\,C^r_{15} + 72\,C^r_{17} + 96\,
%\nonumber\\&&
         C^r_{18} + 44\,C^r_{34} + 48\,C^r_{35}
\nonumber\\&&
 + 8\,C^r_{63}
 + 2\,C^r_{64} + 2\,C^r_{65} + 6\,C^r_{90} )
\nonumber\\&&
       + m_\pi^4\,(  - 30\,C^r_{12} - 16\,C^r_{13} - 12\,C^r_{14} - 8\,C^r_{15} - 40\,C^r_{17} - 48\,
%\nonumber\\&&
         C^r_{18} - 26\,C^r_{34} - 24\,C^r_{35}
\nonumber\\&&
 - 6\,C^r_{63}
 - 4\,C^r_{64} - 2\,C^r_{65} - 5\,C^r_{90} )\,,
\nonumber\\
f_-^E(t) &=& 0\,,
\nonumber\\
f_-^D(t) &=&
       + m_\sigma^2\,t\,( 8\,C^r_{12} - 4\,C^r_{88} + 4\,C^r_{90} )
%\nonumber\\&&
       + m_\pi^2\,t\,(  - 8\,C^r_{12} + 4\,C^r_{88} - 4\,C^r_{90} )
\nonumber\\&&
       + m_\sigma^4\,(  - 24\,C^r_{12} - 16\,C^r_{13} - 8\,C^r_{15} - 16\,C^r_{17} - 8\,C^r_{34} - 4\,
%\nonumber\\&&
         C^r_{64} - 4\,C^r_{65} - 4\,C^r_{90} )
\nonumber\\&&
       + m_\pi^2\,m_\sigma^2\,(  - 16\,C^r_{13} - 16\,C^r_{14} - 8\,C^r_{15} + 16\,C^r_{17} - 8\,C^r_{63} -
%\nonumber\\&&
         4\,C^r_{64} + 4\,C^r_{65} )
\nonumber\\&&
       + m_\pi^4\,( 24\,C^r_{12} + 32\,C^r_{13} + 16\,C^r_{14} + 16\,C^r_{15} + 8\,C^r_{34} + 8\,C^r_{63} +
%\nonumber\\&&
         8\,C^r_{64} + 4\,C^r_{90} )\,.
\ea

\end{document}